\documentclass[a4paper,11pt]{article}
\pdfoutput=1 

\usepackage{jheppub} 
\usepackage{bbm,bm,graphicx,mathtools,color,slashed,hyperref}
\usepackage{amssymb,amsmath}

\graphicspath{{./Figures/}}

\newcommand{\diff}{\mathrm{d}}

\newcommand{\ve}{\varepsilon}
\newcommand{\Diff}{{\mathcal{D}}}

\newcommand{\im}{\mathrm{i}}

\newcommand{\calB}{\mathcal{B}}

\newcommand{\rmd}{\mathrm{d}}
\newcommand{\rme}{\mathrm{e}}

\newcommand{\rmB}{\mathrm{B}}

\newcommand{\sign}{\mathrm{sign}}

\usepackage{mathabx}

\DeclareMathOperator{\tr}{tr}

\preprint{YITP-22-68}

\title{Winding $\theta$ and destructive interference of instantons}

\author[1]{Mendel Nguyen,}
\affiliation[1]{Department of Physics, North Carolina State University, Raleigh, NC 27607, USA}

\emailAdd{mendelnguyen@gmail.com}

\author[2]{Yuya Tanizaki,}
\affiliation[2]{Yukawa Institute for Theoretical Physics,  Kyoto University, Kyoto 606-8502, Japan}

\emailAdd{yuya.tanizaki@yukawa.kyoto-u.ac.jp}

\author[1]{Mithat \"{U}nsal}

\emailAdd{unsal.mithat@gmail.com}

\abstract{
While the $\theta$ dependence of field theories is $2\pi$ periodic, the ground-state wavefunctions at $\theta$ and $\theta+2\pi$ often belong to different classes of symmetry-protected topological states. 
When this is the case, a continuous change of the $\theta$ parameter can introduce an interface that supports a nontrivial field theory localized on the wall. 
We consider the $2$d $\mathbb{C}P^{N-1}$ sigma model as an example and construct a weak-coupling setup of this interface theory by considering the small $S^1$ compactification with nonzero winding $\theta$ parameter and a suitable symmetry-twisted boundary condition. 
This system has $N$ classical vacua connected by fractional instantons, but the anomaly constraint tells us that the fractional-instanton amplitudes should vanish completely to have $N$-fold degeneracy at the quantum level. 
We show how this happens in this purely bosonic system, uncovering that the integration over the zero modes annihilates the fractional instanton amplitudes, in sharp contrast to what happens when the $\theta$ angle is constant.  
Moreover, we provide another explanation of this selection rule by showing that the $N$ perturbative vacua acquire different charges under the global symmetry with the activation of the winding $\theta$ angle.  
We also demonstrate a similar destructive interference between instanton effects in the $\mathbb{C}P^{N-1}$ quantum mechanics with the Berry phase. 
}

\begin{document}
\maketitle

\section{Introduction}

When a continuous field in a quantum field theory (QFT) is classified by a topological charge $Q_{\mathrm{top}}\in \mathbb{Z}$, the QFT has a parameter $\theta$ known as the vacuum angle~\cite{Belavin:1975fg, Callan:1976je, Jackiw:1976pf, tHooft:1976rip}. 
Its interpretation as an angle is due to the fact that it enters the path integral weight through the factor $\exp(\im\, \theta\, Q_{\mathrm{top}})$.
Although the $2\pi$-periodicity in $\theta$ is a rigorous property of the partition function and local correlation functions in closed spacetimes, it is not necessarily a property of the vacuum wavefunction itself. 
As the $2\pi$-periodicity of the partition function only implies the unitary equivalence of the systems at $\theta$ and $\theta+2\pi$, it is possible that the ground-state wave functions at $\theta$ and $\theta+2\pi$ are orthogonal to each other. 
If this is the case, then even if the system is trivially gapped at generic values of $\theta$, there has to be a phase transition as we continuously rotate $\theta$ by $2\pi$, as in fact happens for $4$d Yang-Mills theory~\cite{Witten:1980sp,DiVecchia:1980yfw, Witten:1998uka, tHooft:1981bkw} and the $2$d $\mathbb{C}P^1$ sigma model~\cite{Haldane:1982rj,Haldane:1983ru}. 

Recent advances in generalized symmetries and anomalies provide new perspectives into QFT dynamics and, in particular, clarify the kinematical origin for the existence of the phase transition at $\theta=\pi$. 
In both $4$d Yang-Mills theory and the $2$d $\mathbb{C}P^{N-1}$ sigma model, the partition function with background gauge fields for symmetries is not invariant under a $2\pi$ shift in $\theta$, but acquires a local counter term in the background gauge fields~\cite{Gaiotto:2017yup, Tanizaki:2017bam, Komargodski:2017dmc, Kikuchi:2017pcp, Tanizaki:2018xto, Karasik:2019bxn, Cordova:2019jnf, Cordova:2019uob}. 
This proves that the trivially gapped states at $\theta$ and $\theta+2\pi$ (for generic values of $\theta$) are different symmetry-protected topological (SPT) states, and thus there must be a phase transition separating them. 

When two states belong to different SPT phases, there exists nontrivial physics at the interface between them. 
We can create such an interface by promoting the $\theta$ angle to a position-dependent background field $\theta(x)$ such that $\theta(x\to -\infty)<\pi$ and $\theta(x\to +\infty)>\pi$. 
Since we can regard the interface as the boundary of an SPT state, the dynamics at the interface is subject to an 't~Hooft anomaly matching constraint, which rules out the trivially gapped phase. 
One of the purposes of this paper is to construct a strategy to study its dynamics using reliable semiclassical computations. 

In this paper, we study the $2$d $\mathbb{C}P^{N-1}$ sigma model on $S^1\times \mathbb{R}$ with a position-dependent $\theta$ angle as an emblematic example. 
The $\theta$ angle depends on the compactified direction $x\in S^1$, and moreover it can have a nonzero winding number: 
\begin{align}
    \theta(x+L)=\theta(x)+2\pi w,
    \label{winding1}
\end{align}
with some $w\in \mathbb{Z}$. 
Although such a configuration of $\theta$ seems to have a discrete jump at some location in $S^1$, the angular nature of $\theta$ allows us to make the jump physically transparent~\cite{Cordova:2019jnf, Cordova:2019uob}, as in the cases of symmetry-twisted boundary conditions. 
When the $S^1$ is much larger than the strong-interaction scale, $L\Lambda \gg 1$, the local dynamics should be identical to that of the infinite volume limit. At each place $\theta(x)$ goes across an odd integer multiple of $\pi$, the interface of different SPT states appears and supports the projective representation of $PSU(N)$ flavor symmetry. 
In particular, when $w$ is not a multiple of $N$, the total Hilbert space should belong to a projective representation of the $PSU(N)$ symmetry so that each energy eigenstate must have at least $N$-fold degeneracy, which is a rigorous consequence of the anomaly matching with nonzero winding $\theta$ angle. 

For constant $\theta$, the $PSU(N)$ symmetry-twisted boundary condition provides a suitable framework for the semiclassical analysis of the ground-state property of $\mathbb{C}P^{N-1}$ model~\cite{Dunne:2012ae,Dunne:2012zk}.  
There are $N$ classical vacua in the symmetry-twisted boundary condition, and tunneling processes connect them with the fractional topological charge, which naturally explains the $N$-branch structure of the ground states. 
There is a symmetry reason behind this adiabatic continuity, as the symmetry twist retains the 't~Hooft anomaly of $2$d theory under the reduction to quantum mechanics by $S^1$ compactification~\cite{Tanizaki:2017qhf}. 
Furthermore, in the large-$N$ limit, this symmetry twist is the necessary and sufficient condition for the theory to satisfy volume independence \cite{Sulejmanpasic:2016llc}. 
Motivated by these successes, we discuss the $PSU(N)$ twisted boundary condition with a nonzero winding $\theta$ angle. 

The upshot of our work is the following: depending on whether $\theta$ has a nonzero winding number, the role of fractional instanton becomes completely different. 
In both situations, we have $N$ classical vacua, and fractional instanton configurations interpolate between them. 
For constant values of $\theta \neq \pi$, the fractional instantons lift the degeneracy, and we get the unique ground state. 
For winding $\theta$, however, the anomaly tells us that the quantum vacua should have $N$-fold degeneracy. 
How can the instanton amplitudes completely vanish in a purely bosonic system?

\begin{figure}[t]
\vspace{-1.5cm}
\centering
\includegraphics[width = 1.0\textwidth]{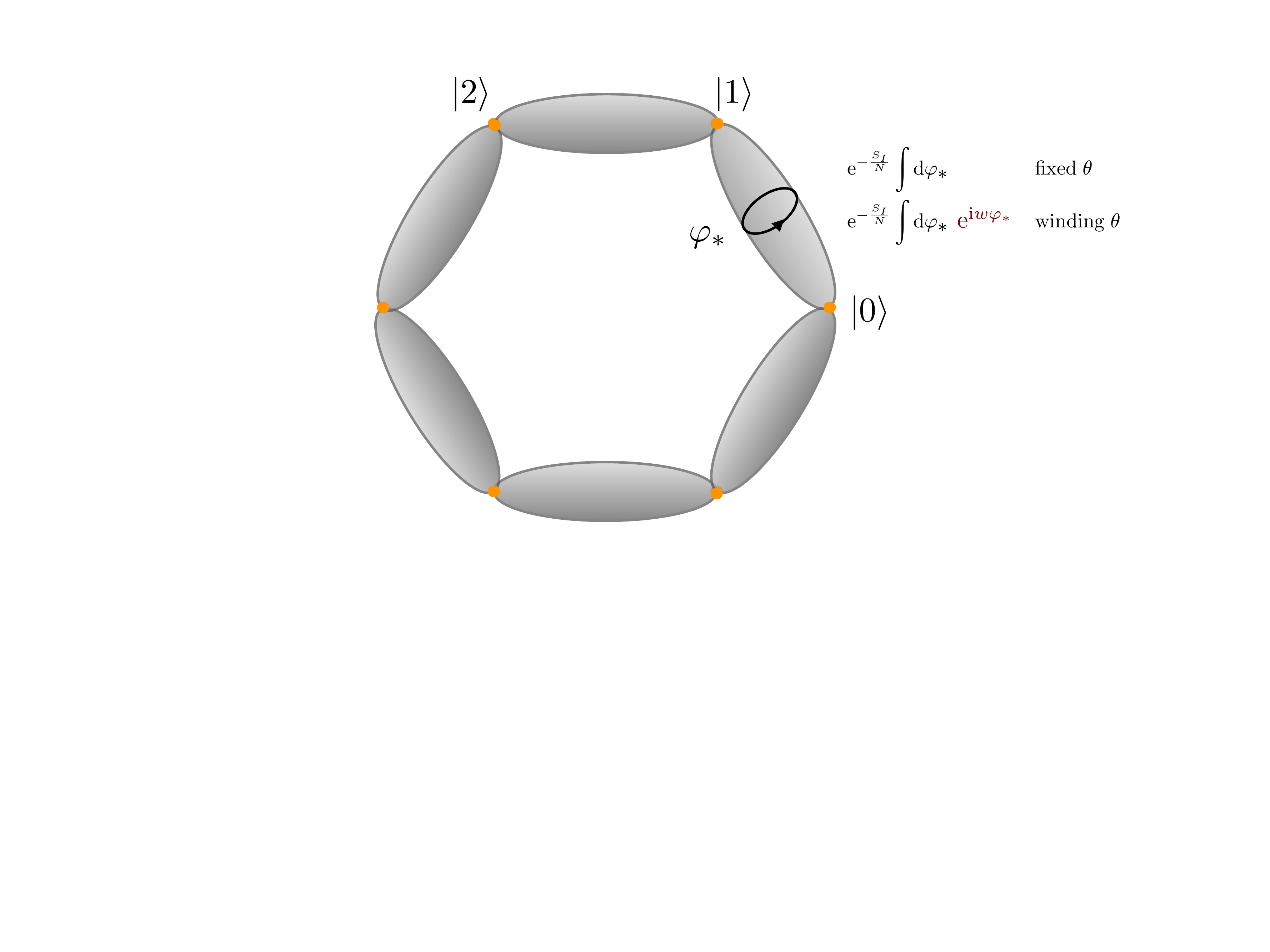}
\vspace{-5.5cm}
\caption{
With symmetry-twisted boundary condition, the $\mathbb CP^{N-1}$  model has $N$ classical vacua and fractional instantons with moduli $\tau_* \in \mathbb R$ and $\varphi_* \in S^1$.  
When $\theta$ has nonzero winding around the spatial circle, the integration over the $\varphi_*$ modulus gives rise to destructive interference which annihilates the transition amplitude.
 }
\label{fig:moduli}
\end{figure}

The dilute  (fractional) instanton gas analysis in a weak coupling regime is valid either with a constant or a nonzero winding $\theta$ angle. Due to subtle effects associated with the definition of such a $\theta$ term, we find that one of the instanton zero-mode directions affects the instanton amplitude and adds a purely imaginary term to the instanton action. 
The integration over that zero-mode direction leads to the vanishing of all transition amplitudes between consecutive classical vacua, i.e. 
\begin{align}
\langle {k+1}| {\rm e}^{- \beta \hat{H}} | k\rangle \sim \left\{ \begin{array}{ll}
 {\rm e}^{-\frac{S_I}{N} + {\rm i} \frac {\theta}{N} } \int _0^{2 \pi} {\rm d} \varphi_* & \qquad {\rm constant}  \;\; \theta , \cr 
  {\rm e}^{-\frac{S_I}{N} + {\rm i} \frac {\bar \theta}{N} } \int _0^{2 \pi} {\rm d} \varphi_* \;    {\rm e}^{ {\rm i} w \varphi_*} = 0  & \qquad {\rm winding}  \;\; \theta ,
\end{array} \right. 
\label{destructive}
\end{align}
where $\varphi_*$ is a bosonic zero mode of the fractional instanton (see Fig.~\ref{fig:moduli}).
The latter gives the destructive interference between Euclidean path histories and the transition amplitude vanishes.  
It is opposite to what happens in the case where $\theta$ is kept fixed as a function of the compact dimension. 
As a consequence, the classical $N$-fold degeneracy is maintained even at the quantum level for the nonzero winding $\theta$ angle. It is important to emphasize that the density of instantons in both Euclidean vacua is the same,  but the effect they lead to are opposite. 

This remarkable effect has an equally interesting implication in the  Hilbert space and operator formalism. The states $|k \rangle$ acquire nontrivial charges of the $U(1)^{N-1}/{\mathbb Z}_N$ symmetry for the winding $\theta$ angle, while they are neutral for constant $\theta$. We show this by simply applying Noether's theorem in the presence of winding theta for a $U(1)$ subgroup.  We find the simple formula
\begin{align}
\hat{Q}| k \rangle  = wk |k \rangle,
\label{eq:statecharging0}
\end{align}
where $w$ is winding numnber \eqref{winding1}.  
This is analogous to the  Witten effect, 
where magnetically charged particles acquire an electric charge once a (constant) theta angle is turned on \cite{ Witten:1979ey}.  
When the states become charged, the vanishing of the transition amplitudes   \eqref{destructive} and the degeneracy of ground states are naturally 
implied by the selection rule.

To elucidate the destructive interference phenomena, we also provide a basic example from quantum mechanics with and without the Berry phase, and this example with the Berry phase also serves as the interface theory between two different SPT states. After studying its 't~Hooft anomaly, we introduce a potential term to have $N$ classical minima and perform the dilute instanton gas approximation. 
Without the Berry phase, these $N$ classical vacua are lifted non-perturbatively due to instanton effects. Once a particular Berry phase is added, various instanton paths between a given pair of degenerate minima add up to zero. As a result of this destructive interference, the $N$-fold degeneracy remains intact, which is quite parallel to the weak-coupling analysis of the nonzero winding $\theta$ angle. 

\section{\texorpdfstring{$2$d}{2d} \texorpdfstring{$\mathbb{C}P^{N-1}$}{CP{N-1}} model and its generalized anomaly}

In this section, we describe the generalized anomaly of $2$d $\mathbb{C}P^{N-1}$ sigma model about its $\theta$ periodicity and discuss its physical consequences. This is basically a brief review of Ref.~\cite{Komargodski:2017dmc}. 
We also give its interpretation from the $SU(N)$ spin chain. 

\subsection{\texorpdfstring{Generalized mixed anomaly between $PSU(N)$}{PSU(N)} symmetry and \texorpdfstring{$\theta$}{theta} periodicity}
\label{sec:anomaly}

The $2$d $\mathbb{C}P^{N-1}$ sigma model is defined by the Euclidean action, 
\begin{equation}
    S=\frac{1}{g^2}\int_{M} |(\diff +\im a)\vec{z}\,|^2+\frac{\im \theta}{2\pi}\int_{M} \diff a. 
\end{equation}
Here, $M$ is the closed $2$d spacetime manifold, $\vec{z}$ is the $\mathbb{C}^N$-valued field with $|\vec{z}\,|^2=1$, and $a$ is the $U(1)$ gauge field. 
In the following sections, we shall extend the case where the $\theta$ parameter depends on the position. Here, for simplicity, let us keep it to be constant. 

The $\theta$ parameter is $2\pi$ periodic due to the Dirac quantization of the $U(1)$ gauge field, and we also have 
\begin{itemize}
    \item $PSU(N)$ symmetry, 
    \item $\mathsf{C}$ symmetry at $\theta\in \pi \mathbb{Z}$. 
\end{itemize}
The $PSU(N)$ symmetry refers to the projective $SU(N)$ rotation of the gauge-invariant spin variables $z^*_i z_j$. 
On the spinon fields $\vec{z}$, it acts as an $SU(N)$ transformation, 
\begin{equation}
    \vec{z}\mapsto U\vec{z}, 
\end{equation}
with $U\in SU(N)$. However, the transformation by center elements, $\rme^{\frac{2\pi\im}{N}k}\bm{1}_N$, can be absorbed by the $U(1)$ gauge redundancy, and the actual symmetry acting on local gauge-invariant operators is given by the quotient group, $PSU(N)\simeq SU(N)/\mathbb{Z}_N$. 

The $\mathsf{C}$ symmetry is the charge conjugation, $z_i\to z^*_i$, which flips the sign of the gauge field $a\to -a$. The kinetic term is manifestly invariant under $\mathsf{C}$, but the $\theta$ parameter effectively flips its sign. 
Therefore, it is explicitly broken at generic values of $\theta$, but it is present when $\theta\in \pi \mathbb{Z}$. 
For $\theta=0$, it is obvious since $\mathsf{C} \colon \theta=0\to -\theta=0$. For $\theta=\pi$, we have $\mathsf{C} \colon \theta=\pi\to -\theta=-\pi \sim \pi$ due to the $2\pi$ periodicity of $\theta$, and we obtain the $\mathsf{C}$ symmetry at $\theta=\pi$. 

In order to detect the 't~Hooft anomaly, let us introduce the $PSU(N)$ background gauge field, which consists of 
\begin{itemize}
    \item $A$: $U(N)$ $1$-form gauge field, 
    \item $B$: $U(1)$ $2$-form gauge field, 
\end{itemize}
satisfying the constraint
\begin{equation}
    NB=\diff \left(\tr(A)\right). 
\end{equation}
The $U(N)$ $0$-form and $U(1)$ $1$-form gauge transformations are given by 
\begin{align}
    &A\mapsto h^\dagger A h-\im h^\dagger \diff h+\lambda \bm{1}_N,\\
    &B\mapsto B+\diff \lambda,
\end{align}
where $h$ is a $U(N)$-valued function and $\lambda$ is a $U(1)$ $1$-form gauge field. We have to keep this invariance in the minimal coupling procedure. 
In particular, invariance of the gauged kinetic term, 
\begin{equation}
    \frac{1}{g^2}|(\diff+\im a+\im A)\vec{z}\,|^2, 
\end{equation}
shows that the dynamical fields $z$ and $a$ should transform as 
\begin{equation}
    \vec{z}\mapsto h^\dagger \vec{z},\quad a\mapsto a-\lambda. 
\end{equation}
As a result, the $U(1)$ field strength $\diff a$ is no longer gauge invariant, so it should be replaced by the gauge-invariant combination $\diff a+B$. 

The partition function with the $PSU(N)$ background gauge field is given by 
\begin{equation}
    Z_\theta[A,B]=\int \Diff \vec{z}^{\,*}\Diff \vec{z} \,\Diff a \exp\left(-\frac{1}{g^2}\int_{M_2}|(\diff+\im a +\im A)\vec{z}\,|^2-\im \frac{\theta}{2\pi}\int_{M_2}(\diff a+B)\right). 
\end{equation}
This violates the $2\pi$ periodic property of the $\theta$ angle very mildly, and we find 
\begin{equation}
    Z_{\theta+2\pi}[A,B]=\exp\left(-\im \int_{M_2}B\right)Z_{\theta}[A,B]. 
    \label{eq:generalized_anomaly}
\end{equation}
That is, the local counterterm of the background gauge field is shifted by the $2\pi$ rotation of $\theta$. 
We note that there do not exist any $2$d local gauge-invariant counterterms that eliminate it. 
If we regard the above $2$d system as a boundary theory of a $3$d bulk topological theory, then the partition function of the combined system,
\begin{equation}
    Z_{\theta}[A,B]\exp\left(\im \int_{M_3}\frac{1}{2\pi}\theta\, \diff B\right),
    \label{eq:3dSPTaction}
\end{equation}
with $\partial M_3=M_2$ satisfies both the gauge invariance and the $2\pi$ periodicity of $\theta$~\cite{Komargodski:2017dmc, Kikuchi:2017pcp} and this is an analogue of the anomaly cancellation via the inflow mechanism. 

This already gives an important constraint: The ground states cannot be trivially gapped at least for one of the $\theta$ angle~\cite{Karasik:2019bxn, Cordova:2019jnf, Cordova:2019uob}. 
If the system is trivially gapped, i.e. has the unique and gapped ground state, then the system belongs to a certain class of $2$d SPT state with $PSU(N)$ symmetry. 
As we can see from $H_2(\mathcal{B}PSU(N), \mathbb{Z})\simeq \mathbb{Z}_N$, there are $N$ different SPT states described by the classical topological action, $k\int_{M_2} B$, with $k\sim k+N$, and thus 
\begin{equation}
    \frac{Z_\theta[A,B]}{|Z_{\theta}[A,B]|}=\exp\left(\im k \int B\right)  
\end{equation}
for a certain value of $k\in \mathbb{Z}_N$. The above relation~\eqref{eq:generalized_anomaly} tells that 
\begin{equation}
    \frac{Z_{\theta+2\pi}[A,B]}{|Z_{\theta+2\pi}[A,B]|}=\frac{\exp(-\im\int_{M_2}B)Z_\theta[A,B]}{|Z_{\theta}[A,B]|}=\exp\left(\im (k-1)\int_{M_2}B\right). 
\end{equation}
Therefore, the level of the SPT action must be shifted under the $2\pi$ rotation of $\theta$. 
As $k$ is a discrete variable, it cannot jump as a function of $\theta$ unless there exists a quantum phase transition. 
Therefore, for a certain value of $\theta=\theta_*\in \mathbb{R}/2\pi \mathbb{Z}$, the system should have a massless excitation or degenerate ground states. 

The presence of the charge-conjugation symmetry at $\theta=0$ and $\theta=\pi$ gives more detailed data. 
We can readily obtain the following:
\begin{align}
    &\mathsf{C}:Z_{\theta=0}[A,B]\mapsto Z_{\theta=0}[A,B],
    \label{eq:C0}\\
    &\mathsf{C}:Z_{\theta=\pi}[A,B]\mapsto Z_{\theta=-\pi}[A,B]=\exp\left(\im\int_{M_2}B\right)Z_{\theta=\pi}[A,B]. \label{eq:Cpi}
\end{align}
While the partition function at $\theta=0$ is manifestly $\mathsf{C}$-symmetric even with the presence of the $PSU(N)$ background gauge field, the partition function at $\theta=\pi$ has the shift of the local counter term. 

Let us assume that the system at $\theta=\pi$ is trivially gapped. If so, the phase of its partition function is given by the SPT action and thus 
\begin{equation}
    Z_{\pi}[A,B]=|Z_{\pi}[A,B]|\exp\left(\im \ell \int B\right)
\end{equation}
with some integer $\ell\sim \ell+N$. Then, \eqref{eq:Cpi} requires that 
\begin{equation}
    -\ell=1+\ell \bmod N. 
\end{equation}
Since this becomes $2\ell=-1 \bmod N$, such an integer $\ell$ does not exist when $N$ is even. 
Equivalently, we can argue that there is no local counterterm that eliminates the anomalous phase of \eqref{eq:Cpi} and thus the anomaly matching condition rules out the trivially gapped phase at $\theta=\pi$. 

When $N$ is odd, we can find the solution, $\ell=\frac{N-1}{2}$. 
Of course, this does not necessarily mean that the phase at $\theta=\pi$ has to be trivially gapped, and it is more natural to assume the spontaneous breaking of $C$ symmetry at $\theta=\pi$ as in the case of even $N$. 
If we assume that the system at $\theta=\pi$ become trivially gapped, it should be described by the specific level of the SPT action, $k=\ell=\frac{N-1}{2}$. 
We can look at the charge-conjugation symmetry at $\theta=0$ given in \eqref{eq:C0}, which shows that the SPT state at $\theta=0$ must have the level $k=0$. 
As these levels are different, the phase transition should separate them. 
Similarly, the SPT action at $\theta=2\pi$ should have the level $k=-1$, so it should also be separated by the SPT state at $\theta=\pi$ by the phase transition. 
The discussion here is the summary of the global inconsistency condition~\cite{Tanizaki:2017bam, Komargodski:2017dmc, Kikuchi:2017pcp, Tanizaki:2018xto}. 
In the large-$N$ limit, the $\mathbb{C}P^{N-1}$ sigma model can be solved exactly as the mean-field analysis becomes exact, and it shows that the system is gapped with the unique ground states for $\theta\not=\pi$~\cite{DAdda:1978vbw}, and that there are doubly degenerate ground states at $\theta=\pi$ due to the spontaneous $\mathsf{C}$ breaking. 
This is believed to be true for $N\ge 3$, and $N=2$ is special because the $\mathbb{C}P^{1}$ sigma model at $\theta=\pi$ describes the $SU(2)$ antiferromagnetic spin chain of half-integer spins~\cite{Haldane:1982rj,Haldane:1983ru}, which is widely believed to have critical behaviors described by the $SU(2)$ level-$1$ WZW theory with the marginally irrelevant $J\bar{J}$ deformations. 
We then assume in the following discussion that the systems at $\theta\not=\pi$ are trivially gapped.

\subsection{\texorpdfstring{$\theta$}{theta} domain wall and interpretation from the \texorpdfstring{$SU(N)$}{SU(N)} spin chain}

So far, the $\theta$ parameter is taken to be a constant in the spacetime. 
Let us extend our discussion to the case when the $\theta$ parameter depends on the spacetime coordinate $(x,\tau)$. 
For simplicity of the discussion, we may take our spacetime as $\mathbb{R}^2$, and $\theta$ depends only on the spatial coordinate $x$. 
As an example, we consider the following profile of $\theta(x)$:
\begin{align}
    \theta(x)=\pi +\delta \theta\, \tanh(x/\ell), 
    \label{eq:thetax}
\end{align}
where $\delta \theta$ is a positive parameter, which is not too large, and $\ell$ characterizes the length scale for the change of $\theta$ near the origin.

When $x\ll -\ell$, $\theta(x)\simeq \pi-\delta \theta$ behaves almost as the constant, and thus the bulk physics should be described by the unique and gapped ground state. 
The same is true for the region $x\gg \ell$, where $\theta(x)\simeq \pi+\delta \theta$ is again almost a constant. 
Then, the question is whether we have an interesting low-energy dynamics around the $\theta$ interface at the origin. 
We will see that there has to be the $N$-fold degeneracy for the effective quantum mechanics localized on the $\theta$ domain wall. 

\begin{figure}[t]
\vspace{-1.5cm}
\centering
\hspace{-0.3cm}
\includegraphics[width = 1.0\textwidth]{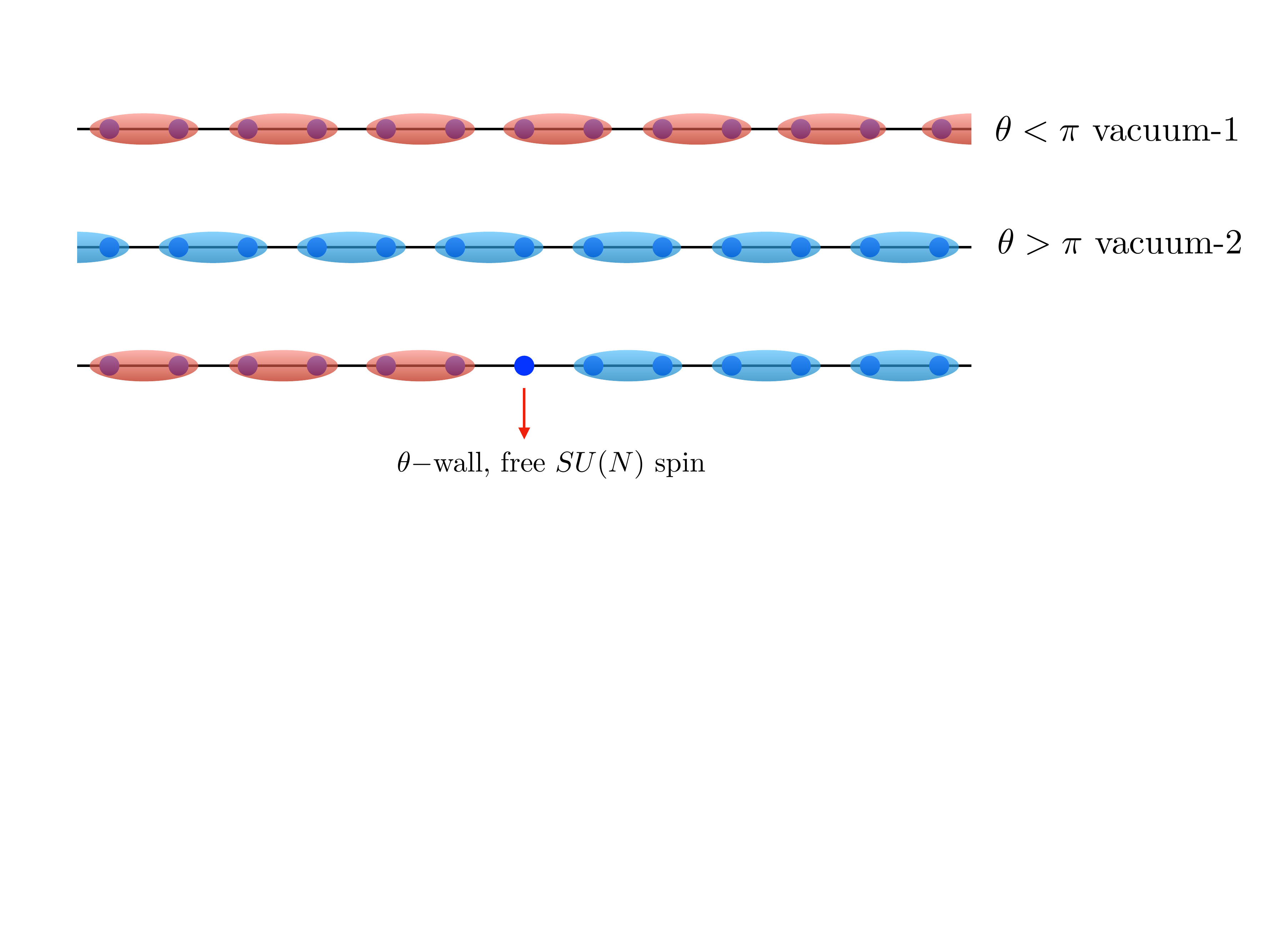}
\vspace{-5.5cm}
\caption{By tuning staggering interaction in an $SU(N)$ spin chain,  one can emulate 
$\theta<\pi$ or $\theta>\pi$, corresponding to two distinct vacua. 
A free $SU(N)$ spin  quantum mechanics lives on the $\theta$-domain wall
 }
\label{fig:wall}
\end{figure}

To get physical intuition, it is convenient to think this problem in the language of the $SU(N)$ spin chain. 
For this purpose, we note that the $2$d $\mathbb{C}P^{N-1}$ sigma model at $\theta=\pi$ can be regarded as the low-energy effective theory of a certain $SU(N)$ spin chain with the fundamental representation on odd sites and the anti-fundamental representation on even sites~\cite{Brower:2003vy, Beard:2004jr, Caspar:2022llo} (we also give the derivation in Appendix~\ref{sec:CPsigma_SUNchain}). 
As the one-unit lattice translation associated with the complex conjugation of $SU(N)$ spins leads to the $\mathbb{Z}_2$ charge conjugation symmetry in the effective field theory, we can slightly shift the $\theta$ parameter from $\pi$ by adding the small staggered interaction that explicitly breaks the one-unit lattice translation. 

We can now understand the two-fold degenerate ground states at $\theta=\pi$ as follows. 
As the two-fold degeneracy is associated with the spontaneous $C$ breaking in $\mathbb{C}P^{N-1}$ sigma model, the lattice translation symmetry $\mathbb{Z}$ should be spontaneously broken to the even-site translations $2\mathbb{Z}$ in the $SU(N)$ spin chain. 
Then, the valence bond solid (VBS) states arise as a natural picture to interpret such ground states. 
That is, the ground-state wave function can be understood as the spin singlet pair between neighboring $SU(N)$ spins. 
The two-fold degeneracy is then the consequence of the fact that the spin singlet can be formed with the equal possibility between odd-even pairs and even-odd pairs (denoted with orange and blue pairs in Fig.~\ref{fig:wall}). 
By adding the staggered interaction, one of the spin singlet pairs is preferred compared with the another and we get the unique ground state.

The space-dependent $\theta$ given in \eqref{eq:thetax} can be realized by flipping the sign of the staggered interaction around $x=0$. 
As shown in Fig.~\ref{fig:wall}, one of the spin at the interface is free from the singlet pairing. 
Therefore, we should have $N$ degenerate states on the wall. 
As the interface state belongs to the projective representation of $PSU(N)$, it cannot be screened by the interaction with the bulk excitations. 

In other words, the ground state for $x<0$ belongs to the level-$0$ SPT state, while the one for $x>0$ belongs to the level-$1$ SPT state with $PSU(N)$ symmetry, as we have discussed. 
Then, the interface at $x=0$ can be regarded as the boundary of the level-$1$ SPT state, so it must support the low-energy excitations in the projective representation of $PSU(N)$ symmetry by the anomaly-inflow mechanism. 

\subsection{Detecting the anomaly with a spatially winding \texorpdfstring{$\theta$}{theta} configuration}
\label{sec:anomaly_windingtheta}

Let us consider the cylinder geometry, $M_2=S^1\times \mathbb{R}$, and we introduces the space-dependent $\theta$ angle, $\theta(x)$ with $x\in S^1$. 
Moreover, we introduce the nonzero winding number, $\theta(x+L)=\theta(x)+2\pi w$ with $w\in \mathbb{Z}$. 
As we will review in Appendix~\ref{sec:spcetimedependent_theta}, the $2\pi$ jump of the $\theta$ parameter can be made physically transparent thanks to the periodic property. 
Therefore, when the size of $S^1$ is large, this is the setup with $w$ interfaces, and each interface supports the fundamental representation of $SU(N)$ as discussed above. 
Then, the Young diagram for the $SU(N)$ representation of the ground states must be given by some $w$-box representation ($\bmod\, N$). 
That is, the Hilbert space of the spatially winding $\theta$ angle has to be in the projective representation of the $PSU(N)$ flavor symmetry when $w\not \in N\mathbb{Z}$.

We can explain this physical intuition in more formal language. 
As we have discussed in \eqref{eq:3dSPTaction}, the $2$d $\mathbb{C}P^{N-1}$ sigma model with the spatially varying $\theta$ angle can be understood as the boundary of the $3$d SPT action, $\frac{1}{2\pi}\int_{M_3} \theta\, \diff B\sim -\frac{1}{2\pi}\int_{M_3}\diff \theta \wedge B$. 
By taking $M_3=S^1\times M'_2$, where $\partial M'_2=\mathbb{R}$ (here we formally regard $\mathbb{R}$ is the infinitely large $S^1$), 
\begin{align}
    \frac{1}{2\pi}\int_{S^1\times M'_2}\diff \theta\wedge B&= \frac{1}{2\pi}\int_{S^1}\diff \theta \int_{M'_2}B\nonumber\\
    &=w\int_{M'_2}B. 
\end{align}
Therefore, the effective quantum mechanics should be regarded as the boundary of the classical $2$d Dijkgraaf-Witten-type theory, and via the anomaly inflow, its Hilbert space is in some projective representation of $PSU(N)$ symmetry with $N$-ality $w \bmod N$.

We note that the above analysis only specifies the $N$-ality of the projective representation, and we need more detailed data to specify exactly which representation appears. 

\section{\texorpdfstring{$\mathbb{C}P^{N-1}$}{CP{N-1}} quantum mechanics with Berry phase}
\label{sec:CPN_QM_Berry}

In this section, we discuss the quantum mechanics with $\mathbb{C}P^{N-1}$ target space with the Berry phase. 
For a suitable choice of the level for the Berry phase, this is expected to be the interface theory. 
Furthermore, this gives a nice exercise for the semiclassical study of nontrivial winding $\theta$ studied in the next section. 

The Euclidean action is 
\begin{equation}
    S=p\int \vec{z}\,^\dagger\partial_\tau \vec{z}\, \diff \tau+\int\left\{ \frac{1}{g^2}|(\partial_\tau+\im a_\tau) \vec{z}\,|^2+V(z^*_i z_j)\right\}\diff \tau. 
\end{equation}
The first term is the Berry phase and the second term describes the kinetic and potential terms. 
This action has the local $U(1)$ gauge invariance, $\vec{z}(\tau)\mapsto \rme^{-\im \alpha(\tau)}\vec{z}(\tau)$ and $a\to a+\diff \alpha$. The gauge invariance of the kinetic term and the potential term is manifest. On the other hand, the Berry phase does not have such a manifest invariance as the time derivative is not gauge covariant. The change of the Berry phase is given by 
\begin{equation}
    S_\rmB=p\int \vec{z}\,^\dagger \diff \vec{z}\mapsto p \int (\rme^{\im \alpha}\vec{z}\,^\dagger) \diff (\rme^{-\im \alpha}\vec{z}\,)=S_\rmB-\im p\int \diff \alpha. 
\end{equation}
Since $\int_{S^1}\diff \alpha\in 2\pi \mathbb{Z}$, the path-integral weight does not change under the $U(1)$ gauge transformation if and only if $p\in \mathbb{Z}$. 
We note that the Berry-phase term naturally appears when we consider the coherent-state path integral of $SU(N)$ spins (see Appendix~\ref{sec:BerryPhase}), and $p\in \mathbb{Z}$ corresponds to the number of boxes in the Young tableaux for totally symmetric representation. 

\subsection{Anomaly of \texorpdfstring{$\mathbb{Z}_N\times \mathbb{Z}_N\subset PSU(N)$}{ZNxZN in PSU(N)}}

\subsubsection*{$PSU(N)$ anomaly}

When the potential term $V$ is absent, this system has $PSU(N)=SU(N)/\mathbb{Z}_N$ symmetry. 
We can introduce the $PSU(N)$ background gauge field $(A,B)$ as we have discussed in Sec.~\ref{sec:anomaly}. 
Let us discuss the Berry phase with the presence of $(A,B)$. In order to have the $U(N)$ gauge invariance, we must replace it by 
\begin{equation}
    S_\rmB=p \int \vec{z}\,^\dagger \diff \vec{z} \,\Rightarrow \, S_\rmB[A]=p\int \vec{z}\,^\dagger (\diff + \im A) \vec{z}. 
\end{equation}
Even though this has the local $U(N)$ gauge invariance, it changes under the $U(1)$ $1$-form gauge transformation, 
\begin{equation}
    S_\rmB[A+\lambda \bm{1}_N]=S_{\rmB}[A]+\im p\int \lambda. 
    \label{eq:PSUN_anomaly}
\end{equation}
To cancel this anomaly, we can introduce the local counterterm, $-\im\frac{p}{N}\int \tr(A)$, but this is not large $U(N)$ gauge invariant unless $p$ is an integer multiple of $N$. 
Therefore, for $p\not \in N\mathbb{Z}$, the relation~\eqref{eq:PSUN_anomaly} describes an 't~Hooft anomaly of $PSU(N)$ symmetry. 
Indeed, this quantum mechanical system can be regarded as the boundary excitations of $(1+1)$d symmetry-protected topological (SPT) system, and its topological action is given by 
\begin{equation}
    S_{2\rmd\, \mathrm{SPT}}[A,B]=\im p \int B, 
    \label{eq:2dSPT}
\end{equation}
with $p\sim p+N$. The anomaly \eqref{eq:PSUN_anomaly} is now canceled by the anomaly-inflow mechanism. 
As a consequence of the anomaly matching, energy eigenstates must have nontrivial degeneracy for $p\not=0 \bmod N$. Especially when $\gcd(p,N)=1$, each energy eigenvalue must have at least $N$ degenerate states. 

\subsubsection*{$\mathbb{Z}_N\times \mathbb{Z}_N$ anomaly}

Let us turn on the potential term $V$, which breaks the $PSU(N)$ spin rotation symmetry to a smaller subgroup. 
We are interested in the case when $V$ preserves the nontrivial subgroup, 
\begin{equation}
    (\mathbb{Z}_N)_{\mathrm{clock}}\times (\mathbb{Z}_N)_{\mathrm{shift}}\subset PSU(N),
\end{equation}
of the spin rotation symmetry. 
We shall see that this carries the essential information of the above 't~Hooft anomaly~\cite{Tanizaki:2018xto}. 

These $\mathbb{Z}_N$ symmetries are generated by the clock and shift matrices in $U(N)$, 
\begin{equation}
    (C)_{mn}=\rme^{\frac{2\pi\im}{N}(n-1)}\delta_{m\,n}, \quad
    (S)_{mn}=\delta_{m+1\,n}, 
    \label{eq:clock_shift}
\end{equation}
where indices are understood in mod $N$. This satisfies 
\begin{equation}
    C^N=S^N=\bm{1}_N, 
\end{equation}
and they commute up to the center elements,
\begin{equation}
    SC=\rme^{\frac{2\pi\im}{N}}CS. 
    \label{eq:algebra_clockshift}
\end{equation}
This shows that the $\mathbb{Z}_N\times \mathbb{Z}_N$ symmetry is projectively realized for the spinon fields $z$. 

In order to see the 't~Hooft anomaly, let us introduce the background gauge field $A_C$ for $(\mathbb{Z}_N)_{\mathrm{clock}}$. It can be realized as the $U(1)$ gauge field with the constraint, 
\begin{equation}
    NA_C=\diff \Phi_C,
\end{equation}
where $\Phi_C$ is a $2\pi$-periodic scalar field. 
We require the $U(1)$ gauge invariance, 
\begin{align}
A_C\mapsto A_C+\diff \alpha, \,\, 
\Phi_C\mapsto \Phi_C + N\alpha,
\end{align}
and 
\begin{align}
z_n\mapsto \rme^{-\im (n-1)\alpha}z_n. 
\end{align}
When we set $\alpha\in \frac{2\pi}{N}\mathbb{Z}$, this is identical to the global $(\mathbb{Z}_N)_{\mathrm{clock}}$ transformation on the spinon field $z$. 
Since $V$ maintains $(\mathbb{Z}_N)_{\mathrm{clock}}$, it can be made invariant under the continuous $U(1)$ symmetry by multiplying an appropriate integer power of $\rme^{\im \Phi_C}$ to each term of $V$. 
To satisfy the above $U(1)$ gauge invariance, the kinetic term is replaced as 
\begin{equation}
    \frac{1}{g^2}\sum_{n=1}^{N}|(\diff + \im a+\im (n-1)A_C)z_n|^2. 
    \label{eq:clock_gauged_kinetic}
\end{equation}
In this way, we can obtain the $(\mathbb{Z}_N)_{\mathrm{clock}}$ gauged action. 

Next, let us determine the $(\mathbb{Z}_N)_{\mathrm{shift}}$ transformation under the presence of the background gauge field $A_C$. It is convenient to require that the gauged kinetic term becomes invariant under the transformation, and it should reduce to the original one $z_n\mapsto z_{n+1}$ and $a\mapsto a$ when we turn off $A_C$. 
After some trials, we can find the following transformation satisfies the above requirements, 
\begin{align}
    (\mathbb{Z}_N)_{\mathrm{shift}}:\,& \left\{\begin{array}{cc}
    z_n\mapsto z_{n+1}, & (n=1,\ldots, N-1)\\ \quad z_N\mapsto \rme^{-\im \Phi_C} z_1, & \\
    a\mapsto a+A_C. &
    \end{array}\right.
    \label{eq:shift_Ac}
\end{align}
It is evident that this reduces to the original transformation when we set $A_C=0$, $\Phi_C=0$. 
We can also confirm that \eqref{eq:clock_gauged_kinetic} is invariant under \eqref{eq:shift_Ac} by noting that $NA_C=\diff \Phi_C$. 

So far, we have seen that $(\mathbb{Z}_N)_{\mathrm{clock}}$ can be gauged in a $(\mathbb{Z}_N)_{\mathrm{shift}}$ symmetric way for the kinetic and potential terms. 
We must examine the property of the Berry phase. The $(\mathbb{Z}_N)_{\mathrm{clock}}$ gauge-invariant Berry phase is given by 
\begin{equation}
    S_\rmB[A_C]=p\int \sum_{n=1}^{N} z^*_n (\diff + \im (n-1)A_C) z_n. 
\end{equation}
Under the $(\mathbb{Z}_N)_{\mathrm{shift}}$ transformation, it changes as 
\begin{align}
    (\mathbb{Z}_N)_{\mathrm{shift}}&:\,S_\rmB[A_C]\nonumber\\
    &\mapsto 
    p\int \left(\rme^{\im \Phi_C}z^*_1(\diff+\im (N-1)A_C)(\rme^{-\im \Phi_C}z_1)+\sum_{n=1}^{N-1}z^*_{n+1}(\diff+\im (n-1)A_C)z_{n+1}\right)\nonumber\\
    &=p\int \sum_{n=1}^{N}z^*_n (\diff+\im (n-2)A_C)z_n\nonumber\\
    &= S_\rmB[A_C]-\im p\int A_C. 
\end{align}
Since $\int_{S^1} A_C\in \frac{2\pi}{N}\mathbb{Z}$, this gives a nontrivial anomaly if $p\not=0 \bmod N$ and it corresponds to the $PSU(N)$ anomaly \eqref{eq:PSUN_anomaly}. 
For $\gcd(p,N)=1$, this requires the $N$-fold degeneracy of each energy eigenvalue despite the fact that the $PSU(N)$ symmetry is explicitly broken down to $\mathbb{Z}_N\times \mathbb{Z}_N$.
Indeed, when we simultaneous introduce the background gauge fields $A_C,A_S$ for the clock and shift $\mathbb{Z}_N$ symmetries, the $B$ field of the $PSU(N)$ gauge field can be replaced as  
\begin{equation}
    B=\frac{N}{2\pi}A_C\wedge A_S, 
\end{equation}
and this quantum mechanical system can be regarded as the boundary of the $2$d SPT action~\eqref{eq:2dSPT} by substituting this expression~\cite{Tanizaki:2018xto}. 

\subsection{Semiclassics, destructive interference, and \texorpdfstring{$N$}{N}-fold degeneracy}

In this section, we shall observe the $N$-fold degeneracy for the $p=1$ case in an explicit manner. 
For this purpose, we introduce a specific potential $V$ that breaks $PSU(N)$ to $\mathbb{Z}_N\times \mathbb{Z}_N$, and perform the semiclassical calculations to find the ground states. 

Let us decompose the potential $V$ as 
\begin{equation}
    V=V_0+V_1, 
\end{equation}
where $V_0\gg V_1$, and thus we can restrict the effective degrees of freedom to the classical vacuum of $V_0$. As a specific choice, we can take
\begin{equation}
    V_0=J_0\sum_{n=1}^{N-1} (|z_n|^2-|z_{n+1}|^2)^2, 
\end{equation}
with $J_0\to \infty$. The classical moduli are now restricted to 
\begin{equation}
    |z_1|=|z_2|=\cdots=|z_N|=\frac{1}{\sqrt{N}},
\end{equation}
and it is parametrized as the $(N-1)$-dimensional torus $T^{N-1}$, 
\begin{equation}
    \vec{z}=\frac{1}{\sqrt{N}}\begin{pmatrix}
    1\\
    \rme^{\im \phi_1}\\
    \vdots\\
    \rme^{\im \phi_{N-1}}
    \end{pmatrix}. 
    \label{eq:classical_moduli}
\end{equation}
This makes computations much easier since the target space is significantly simplified from $\mathbb{C}P^{N-1}$ to $T^{N-1}$. 
Here, we completely fix the $U(1)$ gauge redundancy by declaring $z_1>0$. We note that this can be done without introducing any singularities because $|z_1|=\frac{1}{\sqrt{N}}\not =0$ everywhere. 
Substituting this expression~\eqref{eq:classical_moduli} into the action, we obtain 
\begin{equation}
    S=\frac{\im p}{N}\int \sum_{n=1}^{N-1}(\diff \phi_n)+\int \diff \tau\left(\frac{1}{2}\sum_{n,m=1}^{N-1}\dot{\phi}_n M_{nm} \dot{\phi}_m+V_1(\phi_n)\right),
\end{equation}
where $\dot{\phi}_n=\partial_\tau \phi_n$ and  the mass matrix $M_{nm}$ is given by
\begin{equation}
    M_{nm}=\frac{2}{g^2N}\left(\delta_{nm}-\frac{1}{N}\right). 
\end{equation}
Here, we describe $V_1$ as a function of $\phi_n$. 

Let us determine how the $\mathbb{Z}_N\times \mathbb{Z}_N$ symmetry acts on the fields $\phi_n$. The $(\mathbb{Z}_N)_{\mathrm{clock}}$ symmetry is obvious and it acts as 
\begin{equation}
    (\mathbb{Z}_N)_{\mathrm{clock}}: \phi_n\mapsto \phi_n+\frac{2\pi}{N}n.
\end{equation}
The $(\mathbb{Z}_N)_{\mathrm{shit}}$ symmetry is less trivial, and it is good to work on the $z$ field. It turns out that we should combine the $U(1)$ gauge transformation after the multiplication of the shift matrix $S$:
\begin{align}
    \vec{z}=\frac{1}{\sqrt{N}}\begin{pmatrix}
    1\\
    \rme^{\im \phi_1}\\
    \vdots\\
    \rme^{\im \phi_{N-1}}
    \end{pmatrix} 
    &\xrightarrow{(\mathbb{Z}_N)_{\mathrm{shift}}}S\vec{z}=\frac{1}{\sqrt{N}}\begin{pmatrix}
    \rme^{\im \phi_1}\\
    \rme^{\im \phi_2}\\
    \vdots\\
    \rme^{\im \phi_{N-1}}\\
    1
    \end{pmatrix} \nonumber\\
    &\xrightarrow{U(1)_{\mathrm{gauge}}}
    \frac{1}{\sqrt{N}}\begin{pmatrix}
    1\\
    \rme^{\im (\phi_2-\phi_1)}\\
    \vdots\\
    \rme^{\im (\phi_{N-1}-\phi_1)}\\
    \rme^{-\im \phi_1}
    \end{pmatrix} . 
\end{align}
As a result, the $(\mathbb{Z}_N)_{\mathrm{shift}}$ symmetry acts on $\phi_n$ as follows, 
\begin{align}
    \phi_n\mapsto \phi_{n+1}-\phi_1\,\, (n=1,\ldots, N-2), \quad 
    \phi_{N-1}\mapsto -\phi_{1}. 
\end{align}
Although it is not so apparent, it is straightforward to confirm that the kinetic term is invariant under the $(\mathbb{Z}_N)_{\mathrm{shift}}$ symmetry. 
The Berry phase transforms as 
\begin{equation}
    S_\rmB\mapsto \frac{\im p}{N}\int\left(\sum_{n=1}^{N-2}(\diff \phi_{n+1}-\diff \phi_1)-\diff \phi_1\right)=S_\rmB-\im p\int \diff \phi_1,
    \label{eq:Berry_shift}
\end{equation}
and thus the path-integral weight is invariant for $p\in \mathbb{Z}$. 

The potential term $V_1(\phi_n)$ must satisfy the above symmetry requirements. 
For our purpose, its detailed form is not important at all, and let us assume that its classical vacua are given by the following $N$ sets,
\begin{equation}
    \vec{\phi}_{(k)}=\begin{pmatrix}
    \phi_{(k),1}\\
    \phi_{(k),2}\\
    \vdots\\
    \phi_{(k),N-1}
    \end{pmatrix}
    =\frac{2\pi k}{N}
    \begin{pmatrix}
    1\\
    2\\
    \vdots\\
    N-1
    \end{pmatrix},  
\end{equation}
with $k=0,1,\ldots, N-1$. For example, in the case of $N=3$, we can realize this condition by setting 
\begin{align}
    V_1(\phi_1,\phi_2)=&-J_1\left(\cos(3\phi_1)+\cos(3\phi_2)+\cos(3(\phi_1-\phi_2))\right) \notag\\
    &-J_2\left(\cos(\phi_1+\phi_2)+\cos(\phi_1-2\phi_2)+\cos(2\phi_1-\phi_2)\right), 
\end{align}
with $J_1, J_2>0$. 
Each vacuum $\vec{\phi}_{(k)}$ is invariant under $(\mathbb{Z}_N)_{\mathrm{shift}}$, 
\begin{equation}
    \vec{\phi}_{(k)}\xrightarrow{(\mathbb{Z}_N)_{\mathrm{shift}}}
    \frac{2\pi k}{N}
    \begin{pmatrix}
    1\\
    \vdots\\
    N-2\\
    -1
    \end{pmatrix}
    =\vec{\phi}_{(k)}
    -\begin{pmatrix}
    0\\
    \vdots\\
    0\\
    2\pi k
    \end{pmatrix}
    \sim \vec{\phi}_{(k)},
\end{equation}
and the $(\mathbb{Z}_N)_{\mathrm{clock}}$ symmetry permutes them in a cyclic way, $\vec{\phi}_{(k)}\mapsto \vec{\phi}_{(k+1)}$. 

\begin{figure}[t]
\vspace{-1.5cm}
\centering
\hspace{-0.3cm}
\includegraphics[width = 1.0\textwidth]{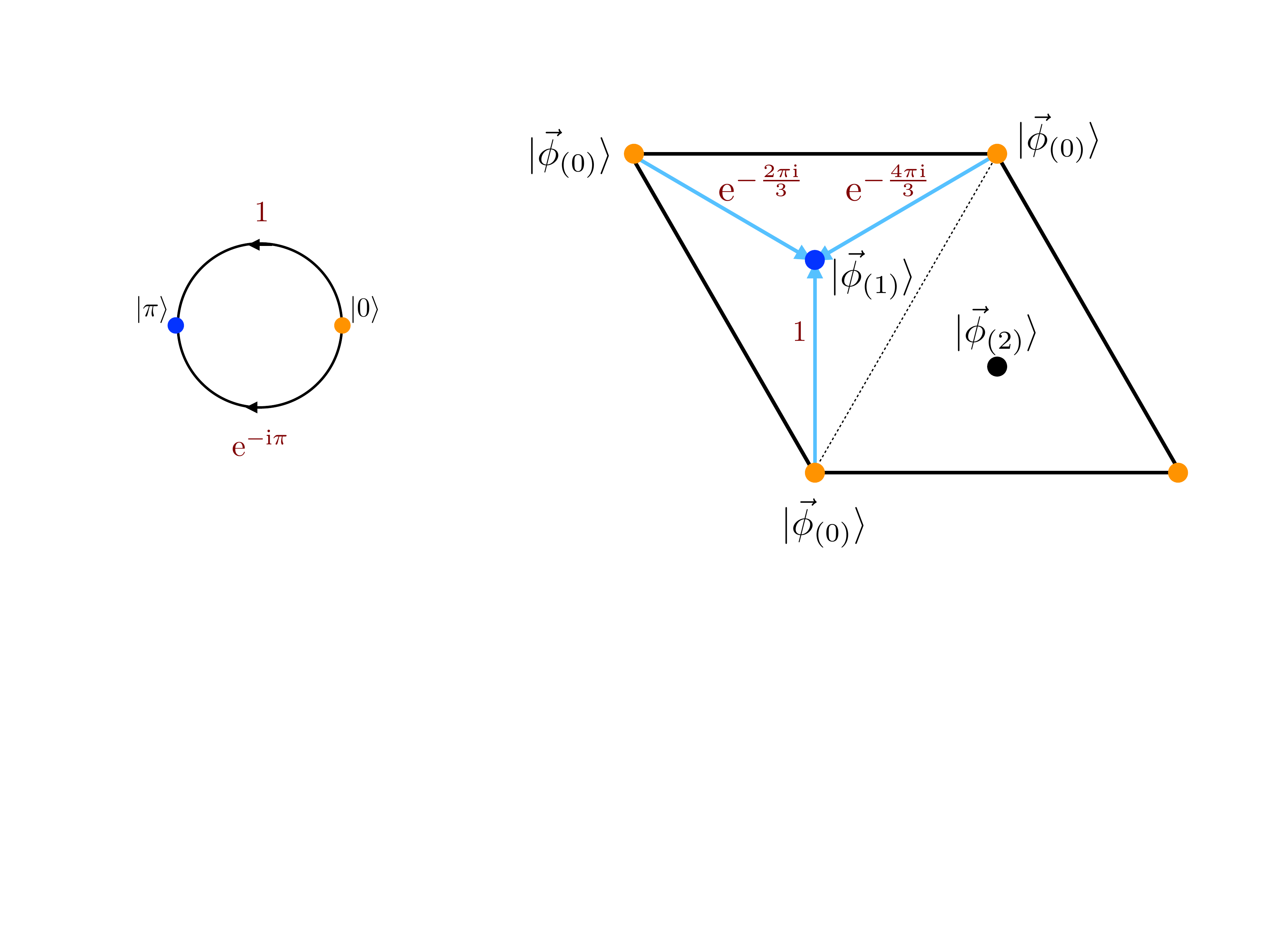}
\vspace{-5.0cm}
\caption{
In the  $ (\mathbb{Z}_N)_{\mathrm{clock}}\times (\mathbb{Z}_N)_{\mathrm{shift}}$ symmetry preserving  perturbation of the  ${\mathbb {CP}}^{N-1}$ models (Figures are for $N=2$ and $N=3$), there exists $N$ isolated vacua. Between a fixed pair of distinct vacua, 
there exists $N$ topologically distinct  tunneling paths. The instanton actions associated with these paths are equal due to symmetries. However, each path is associated with a distinct Berry phase. As a result, there exists an exact destructive interference between tunneling paths, and   the $N$-fold vacuum degeneracy  survives quantum mechanically. This is the semi-classical Berry phase reason behind the anomaly. 
 }
\label{fig:tunneling}
\end{figure}

We would like to discuss if these $N$ classical vacua are lifted by the quantum effect or not. 
For this purpose, let us consider the imaginary-time tunneling process $\vec{\phi}(\tau)$ from $\vec{\phi}_{(0)}$ to $\vec{\phi}_{(k)}$, i.e. $\vec{\phi}(0)=\vec{\phi}_{(0)}$ and $\vec{\phi}(T)=\vec{\phi}_{(k)}$. 
When such a tunneling path exists, there is always another path related by the $(\mathbb{Z}_N)_{\mathrm{shift}}$ symmetry, $S \cdot \vec{\phi}(\tau)$. 
We note that the initial and final points are the same for $\vec{\phi}(\tau)$ and $S \cdot \vec{\phi}(\tau)$, but these paths are topologically distinct in $T^{N-1}$. 
Because of symmetry, the real parts of their Euclidean actions must be the same for those paths. 
However, the Berry phase can be different because of its transformation property~\eqref{eq:Berry_shift}, and we find that 
\begin{align}
    S_\rmB[S \cdot \vec{\phi}(\tau)]&= S_\rmB[\vec{\phi}(\tau)]-\im p \int_{\tau=0}^{\tau=T}\diff \phi_1 \notag\\
    &=S_\rmB[\vec{\phi}(\tau)]-\frac{2\pi \im }{N} kp. 
\end{align}
Therefore, the Berry phase can cause the destructive interference. 
When there is not the Berry phase, i.e. $p=0$, such an interference does not occur and we obtain the unique ground states. 
For general values of $p$, however, we get 
\begin{align}
    \sum_{\ell=0}^{N-1}\rme^{-S_\rmB[S^\ell\cdot \vec{\phi}(\tau)]}=\rme^{-S_\rmB[\vec{\phi}(\tau)]} 
    \left\{\begin{array}{cc}
    =0 &\quad  pk\not \in N\mathbb{Z},\\
    \not=0 &\quad  pk \in N \mathbb{Z}.
    \end{array}
    \right.
\end{align}
The destructive interference occurs unless $k$ is a multiple of $N/\gcd(p,N)$. 
We then obtain $N/\gcd(p,N)$ degenerate vacua. 
Especially when $p=1$, the tunneling effects completely cancel with each other (see Fig.~\ref{fig:tunneling}). As a result, the $N$-fold degeneracy of the ground states is robust, and this is consistent with the anomaly argument.

\section{Semiclassical analysis of the \texorpdfstring{$2$}{2}d \texorpdfstring{$\mathbb{C}P^{N-1}$}{CP{N-1}} model on \texorpdfstring{$S^1\times \mathbb{R}$}{S1xR} with winding \texorpdfstring{$\theta$}{theta}}

In this section, we perform a semiclassical analysis of the $2$d $\mathbb{C}P^{N-1}$ sigma model on $S^1\times \mathbb{R}$.
In order to understand the physics of the $\theta$ interface, we promote $\theta$ to a background field with nonzero winding number $w \in \mathbb{Z}$ around the $S^1$ direction: $\int_{S^1}\diff \theta=2\pi w$. 

Since $\theta$ has nontrivial winding, it must be a nonconstant function on $S^1$. In this case, the proper definition of the $\theta$ term $\frac{1}{2\pi}\int`` \theta\, \diff a "$ requires some care, as $\theta$ is not a genuine real-valued function. For the general definition on arbitrary closed 2d manifolds, see Ref.~\cite{Cordova:2019jnf} or the brief review in Appendix~\ref{sec:spcetimedependent_theta}. 
For our specific situation in the cylinder spacetime, a more naive approach can be given as follows. Let $x$ be the coordinate of $S^1$ with the periodic identification $x \sim x + L$ and regard $\theta$ as an angle-valued function on $[0,L]$. Then by choosing a smooth real-valued lift $\hat{\theta}\colon [0,L] \to \mathbb{R}$ of $\theta$, we might attempt to define 
\begin{equation}
    \frac{1}{2\pi} \int_{S^1 \times \mathbb{R}} `` \theta\, \diff a "
    \overset{?}{  \coloneq } \frac{1}{2 \pi} \int_{[0,L] \times \mathbb{R}} \hat{\theta} \, \diff a.
\end{equation}
This, however, is not quite satisfactory. Because of the nontrivial winding of $\theta$, we must have $\hat{\theta}(L) = \hat{\theta}(0) + 2 \pi w$, and therefore when we identify $x=L$ with $x=0$, $\hat{\theta}$ suffers a $2 \pi w$ jump, which implies a physical defect at $x=0 \sim L$. Indeed, if we try to move the location of the defect from $x=0$ to some other place $x=x_0$, then the value of the above integral is changed by the amount
\begin{align}
    w \int_{-\infty}^{+\infty} a_\tau(x,\tau) \, \diff \tau \,\,\bigg|_{x=0}^{x=x_0}.
\end{align}
To cancel this contribution, we dress the defect with the charge-$w$ temporal Wilson line. That is to say, if we take the definition
\begin{align}
    \frac{1}{2\pi} \int_{S^1 \times \mathbb{R}} `` \theta\, \diff a "
      \coloneq  \frac{1}{2 \pi} \int_{[0,L] \times \mathbb{R}} \hat{\theta} \, \diff a
    - w \int_{-\infty}^{+\infty} a_\tau(0,\tau) \, \diff \tau,
\end{align}
then the location of the $2\pi w$ jump of $\hat{\theta}$  at $x=0$ can be moved to any other place on $S^1$ as long as the Wilson line is moved with it. It is with this definition of the $\theta$ term that we will work,
although in what follows we abuse notation and conflate $\theta$ with its lift $\hat{\theta}$.

If the size of $S^1$ is much larger than the strong scale, $L\Lambda \gg 1$, then we have a $\theta$ interface whenever $\theta$ crosses $\theta=\pi$ which separates different SPT states with $PSU(N)$ symmetry. 
At an interface, there has to be a physical degree of freedom in a projective representation of $PSU(N)$ symmetry, which gives rise to spectral degeneracy on the interface. 
We note that since the $2\pi$ jump of $\theta$ is made physically transparent, there is no anti-wall in this setup. 
In the following, we study the physics when the size of $S^1$ is sufficiently small, $NL\Lambda\lesssim 1$, using semiclassical computations with the flavor-twisted boundary condition. 

\subsection{Flavor-twisted boundary condition}

Some compactifications provide settings in which strong coupling phenomena can be continuously connected to the weak-coupling regimes where one gains control over nonperturbative dynamics \cite{Unsal:2008ch, Unsal:2007jx}. 
In the $2$d $\mathbb{C}P^{N-1}$ sigma model, such a compactification can be achieved by using the flavor-twisted boundary condition
\begin{align}
    \vec{z}(x+L,\tau) = C \cdot \vec{z}(x,\tau),
\end{align}
where $C$ is the clock matrix~\eqref{eq:clock_shift}, or in components,
\begin{equation}
    z_n(x+L,\tau)=\rme^{\frac{2\pi\im}{N}(n-1)}z_n(x,\tau). 
\end{equation}
With this choice of boundary condition, the $2$d $\mathbb{C}P^{N-1}$ sigma model can be studied using reliable semiclassical computations when $NL\Lambda \lesssim 1$, and its dynamics is adiabatically connected to the one in the infinite spacetime $\mathbb{R}^2$ \cite{Dunne:2012ae,Dunne:2012zk, Misumi:2014jua, Misumi:2016fno, Fujimori:2016ljw, Fujimori:2018kqp}. The large-$N$ limit of this set-up satisfies volume independence and  the compactified theory is equivalent to the theory on $\mathbb R^2$ \cite{Sulejmanpasic:2016llc}.  Moreover, the symmetry-twisted boundary condition plays an essential role in maintaining the anomaly-matching constraint on the ground states as clarified in Ref.~\cite{Tanizaki:2017qhf} (see also Refs.~\cite{Shimizu:2017asf, Tanizaki:2017mtm, Dunne:2018hog}).

\subsubsection{Symmetries and anomalies}

\label{sec:FTBC_Anomalies}

When we take the flavor-twisted boundary condition, the continuous part of the $PSU(N)$ flavor symmetry is broken to its maximal Abelian subgroup $U(1)^{N-1}/\mathbb{Z}_N$. 
In addition, there is the $\mathbb{Z}_N$ shift symmetry that cyclically permutes the flavor label combined with the $\mathbb{Z}_N$ center transformation~\cite{Tanizaki:2017qhf, Cherman:2017tey} (see also \cite{Kouno:2012zz, Kouno:2013zr, Kouno:2013mma}). 
This contains the $(\mathbb{Z}_N)_{\mathrm{clock}}\times (\mathbb{Z}_N)_{\mathrm{shift}}$ subgroup, 
and the discussion in Sec.~\ref{sec:anomaly_windingtheta} shows that the effective quantum mechanical system can be regarded as the boundary of the $2$d SPT phase, 
\begin{equation}
    \frac{N w}{2\pi}\int A_C\wedge A_S, 
\end{equation}
where $A_C$ and $A_S$ are the $\mathbb{Z}_N$ gauge fields for the clock and shift symmetries, respectively. 
The minimal number of the ground states to match this anomaly is given by 
\begin{equation}
    \frac{N}{\mathrm{gcd}(N,w)}, 
\end{equation}
and this would be the natural guess for the number of degenerate ground-states.
Although we have restricted our discussion to the $(\mathbb{Z}_N)_{\mathrm{clock}}\times (\mathbb{Z}_N)_{\mathrm{shift}}$ subgroup of the full symmetry group $(U(1)^{N-1}/\mathbb{Z}_N)\rtimes (\mathbb{Z}_N)_{\mathrm{shift}}$, we reach the same conclusion even when we take into account the full symmetry group~\cite{Tanizaki:2018xto, Misumi:2019dwq}. 

\subsubsection{Classical vacua}

A key feature of the flavor-twisted boundary condition is that, at the classical level, there are only $N$ vacua, whereas in the absence of the twist, there is a whole manifold of classical vacua given by $\mathbb{C}P^{N-1}$. To see this, let us look for zero action configurations, which we may take to be independent of the time. Such a configuration $z$ must satisfy the equations
\begin{align*}
    \partial_\tau \vec{z} = 0, \quad \partial_x \vec{z} = -\im a_x \vec{z}.
\end{align*}
It follows that $z$ must take the form
\begin{align*}
    \vec{z}(x,\tau) = \vec{\zeta} \, \rme^{\im \sigma (x)}
\end{align*}
where $\vec{\zeta}$ is a constant unit vector in $\mathbb{C}^N$ and $\partial_x \sigma(x)=-a_x(x)$. By a constant $U(1)$ gauge transformation, we may set $\sigma(0) = 0$. 
Now, we impose the requirement that $\vec{z}$ obey the flavor-twisted boundary condition, and it gives
\begin{align*}
    C \cdot \vec{\zeta} =\vec{\zeta} \, \rme^{\im \sigma(L)}; 
\end{align*}
that is, $\vec{\zeta}$ is an eigenvector of $C$ with eigenvalue $\rme^{\im \sigma(L)}$. This eigenvalue problem is solved by
\begin{align*}
    \vec{\zeta} = \lambda\, \hat{e}_{k+1}, \quad \sigma(L) = \frac{2 \pi k}{N}  \mod 2 \pi, 
\end{align*}
for some integer $k \in \{0,\ldots,N-1\}$ and $\lambda\in \mathbb{C}$ of unit modulus.
Here, $\hat{e}_{k+1}$ is the unit vector along the $(k+1)$-th direction, $\hat{e}_{k+1}=(0,\ldots, 0, 1, 0,\ldots, 0)^\mathrm{T}$. 
By applying a $\tau$-independent $U(1)$ gauge transformation, we can take $\lambda = 1$ and $\sigma(x)$ in the form
\begin{align*}
    \sigma(x) = \frac{2 \pi k x}{NL}.
\end{align*}
Thus, we have found exactly $N$ gauge-inequivalent classical vacua $|k\rangle$, where the wave function of $|k\rangle$ is supported on the gauge equivalence class of the field configuration
\begin{align}
    \vec{\zeta}_{(k)}(x) = \exp\left(\frac{2 \pi \im kx }{NL}\right) \hat{e}_{k+1}.
    \label{eq:classicalvacua}
\end{align}
Using the relation $a = \im \vec{z}\,^\dagger \diff \vec{z}$, we find that the gauge field associated with $\vec{\zeta}_{(k)}$ is given by
\begin{align}
    a_\tau [\vec{\zeta}_{(k)}]=0, \quad a_x[\vec{\zeta}_{(k)}] = -\frac{2 \pi k}{NL},
\end{align}
which is a flat connection with holonomy $\exp(- 2\pi \im k/N)$ around $S^1$. 
Evidently, these classical vacua are invariant under the clock symmetry and cyclically permuted by the shift symmetry. 
By exchanging the role of clock and shift matrices, the situation is quite analogous to the case of $\mathbb{C}P^{N-1}$ quantum mechanics with the Berry phase discussed in Sec.~\ref{sec:CPN_QM_Berry}. 

\subsubsection{Fractional instantons}

Let us discuss the tunneling solution connecting these classical vacua, and another key feature of the flavor-twisted boundary condition is that the tunneling configuration has the fractional topological charge, $\frac{1}{2 \pi} \int_{S^1 \times \mathbb{R}} \diff a \in \frac{1}{N} \mathbb{Z}$. 
Such fractional instantons of $\mathbb{C}P^{N-1}$ sigma model were first discussed in Refs.~\cite{Eto:2004rz, Eto:2006pg, Eto:2006mz} and later used for the semiclassics on $S^1\times \mathbb{R}$~\cite{Dunne:2012ae,Dunne:2012zk}.

To see the fractionalized topological charge, we should note that the requirement of finite action implies that in the far past and future, $\vec{z}$ must approach a classical vacuum configuration. Thus, up to gauge equivalence, we must have
\begin{align}
    \vec{z}(x,\tau \to \pm \infty) = \vec{\zeta}_{(k_\pm)}(x) = \exp\left( \frac{2 \pi \im}{NL}k_\pm x\right) \vec{\zeta}_{(k_\pm)}(0).
\end{align}
Using the fact $a_x = \im \vec{z}\,^\dagger \partial_x \vec{z}$, we find $a_x(\tau \to \pm \infty) = -{2 \pi k_\pm}/{NL}$; whence the topological charge is given by
\begin{align}
    \frac{1}{2 \pi} \int_{S^1 \times \mathbb{R}} \diff a 
    = - \frac{1}{2 \pi} \left. \int_{S^1} a_x \, \diff x \; \right|_{\tau = - \infty}^{\tau = +\infty} = \frac{k_+ - k_-}{N} \in \frac{1}{N}\mathbb{Z}.
\end{align}
This is the $1/N$ quantization, as claimed. 

To find the explicit form of the fractional instanton, we use the Bogomol'nyi-Prasad-Sommerfield (BPS) equality: 
\begin{equation}
    \int_{S^1\times \mathbb{R}} |(\mathrm{D}_\tau\pm \im\, \mathrm{D}_x)\vec{z}\,|^2 \diff^2 x=\int_{S^1\times \mathbb{R}} |\mathrm{D}_\mu \vec{z}\,|^2\diff^2 x 
    \mp \int_{S^1\times \mathbb{R}} \diff a. 
\end{equation}
Here, $\mathrm{D}_\mu=\partial_\mu+\im a_\mu$. For the tunneling from $|0\rangle$ to $|1\rangle$, we set $\vec{z}=\frac{1}{\sqrt{1+|n|^2}}(1,n,0,\ldots, 0)^\mathrm{T}$ with $n\to 0$ as $\tau\to -\infty$ and $|n|\to \infty$ as $\tau\to \infty$. The BPS equation becomes $(\partial_\tau+\im \partial_x)n=0$ and thus $n$ should be a holomorphic function of the complex coordinate $\tau+\im x$. 
The flavor-twisted boundary condition fixes it completely up to gauge transformation and classical moduli, which gives
\begin{align}
    \vec{z}_{(0)}(x,\tau;x_*,\tau_*) = \frac{1}{\sqrt{1 + |\rme^{u-u_*}|^2}} 
    \begin{pmatrix}
    1 \\
    \rme^{u-u_*} \\
    0 \\
    \vdots \\
    0
    \end{pmatrix}
    \label{eq:FracInst}
\end{align} 
Here, we have defined the complex coordinate $u   \coloneq  \frac{2 \pi}{NL}(\tau + \im x)$, and $u_* = \frac{2 \pi}{NL}(\tau_* + \im x_*)$ is a complex collective coordinate with $\tau_* \in \mathbb{R}$ and $x_* \in [0,NL]$. 
The associated $U(1)$ gauge field is given by 
\begin{align}
    a_\tau[\vec{z}_{(0)}] = 0, \quad a_x[\vec{z}_{(0)}] &= -\frac{2 \pi}{NL}\frac{1}{1+ \rme^{- \frac{4 \pi}{NL} (\tau-\tau_*)}}.
\end{align}
Crucially, there is no size modulus, and so the dilute fractional instanton gas approximation does not suffer from infrared divergences.  
The other fractional instanton configurations are given by tunneling between consecutive vacua
\begin{align}
 |0\rangle \rightarrow  |1\rangle \rightarrow  |2\rangle \rightarrow  \cdots  \rightarrow|N-1\rangle \rightarrow  |0\rangle \cdots 
\end{align}
and 
can be obtained by applying the shift symmetry operator. 
Recall that  each fractional instanton configuration has exactly two-zero modes. $N$ of these fractional instantons make the  2d instanton with topological charge one, which has $2N$ bosonic zero modes.  This indeed accounts the number of zero modes of the 2d instanton. 


\subsection{Semiclassical destructive interference with winding \texorpdfstring{$\theta$}{theta}} 
\label{sec:destructive_interference}

Using the fractional instanton solution~\eqref{eq:FracInst}, we can now compute the imaginary-time transition amplitudes
\begin{align}
    \langle k+1| \rme^{-\beta \hat{H}} | k \rangle
\end{align}
in the limit $\beta \to \infty$ within the leading-order semiclassical approximation of the path integral. 
In most cases, these are nonvanishing, and as a result, the degeneracy of the classical vacua is lifted, giving a unique quantum ground state. This is indeed what happens when $\theta$ is a constant generic value. However, when $\theta$ has nontrivial winding $w$, the anomaly discussed in Sec.~\ref{sec:FTBC_Anomalies} requires that at the quantum level, we should have a groundstate degeneracy of at least $N/\mathrm{gcd}(N,w)$. This suggests that the transition amplitudes $\langle k+1| \rme^{-\beta H}|k \rangle$ should completely vanish.
How does this happen?

As we will now show, when $\theta$ has a nonzero winding number, the integration over the compact modulus $x_*$ gives total destructive interference in the transition amplitude. 
At first sight, this may seem impossible, for as we computed above, the gauge configuration $a[\vec{z}_{(0)}]$ of the fractional instanton does not have any $x_*$ dependence, and so neither would the $\theta$ term $\frac{1}{2 \pi} \int ``\theta \, \diff a[\vec{z}_{(0)}]"$. 
The key point is to take the \textit{correct} boundary condition for $\tau\to \pm \infty$, which requires applying a time-dependent gauge transformation that will affect the $\theta$ term. 

Let us look at the fractional-instanton solution~\eqref{eq:FracInst} in more detail. 
Taking the limit $\tau\to -\infty$ and $\tau\to +\infty$, we find 
\begin{align}
    \lim_{\tau\to -\infty}\vec{z}_{(0)}(\tau,x;\tau_*,x_*)=
    \begin{pmatrix}
    1 \\
    0 \\
    0 \\
    \vdots \\
    0
    \end{pmatrix}, \quad 
    \lim_{\tau\to +\infty}\vec{z}_{(0)}(\tau,x;\tau_*,x_*)=
    \begin{pmatrix}
    0 \\
    \rme^{\frac{2\pi \im}{NL}(x-x_*)} \\
    0 \\
    \vdots \\
    0
    \end{pmatrix}. 
\end{align}
We note that the configuration at $\tau=\infty$ has an overall phase depending on $x_*$ and thus it is in a different gauge compared with $\vec{\zeta}_{(1)}$ in \eqref{eq:classicalvacua}. 
This means that $\exp(-S[\vec{z}_{(0)}(\tau,x;\tau_*,x_*)])$ with different $x_*$ compute the transition amplitude from $|0\rangle$ to $|1\rangle$ with different choice of the gauge at the future boundary. 
However, the physical amplitude should be given by a path integral over configurations with fixed boundary conditions at past and future infinity:
\begin{align}
    \lim_{\beta \to \infty} \langle 1| \rme^{-\beta \hat{H}} |0 \rangle
    \sim
    \lim_{\beta \to \infty} \langle \vec{\zeta}_{(1)} | \rme^{-\beta \hat{H}} | \vec{\zeta}_{(0)} \rangle
    = 
    \underset{\begin{subarray}{l}
    \vec{z}(\tau \to -\infty) = \vec{\zeta}_{(0)} \\
    \vec{z}(\tau \to +\infty) = \vec{\zeta}_{(1)}
    \end{subarray}}{\int \Diff \vec{z}^{\,*} \, \Diff \vec{z}}  
    \rme^{- S[\vec{z}\,]} . 
    \label{bcinfty}
\end{align}
Therefore, we should fix the gauge at the boundaries before integrating over the moduli to obtain the physical transition amplitude.

For this purpose, we introduce the $\tau$-dependent gauge transformation, 
\begin{align}
    \vec{z}_{(0)}\mapsto h^\dagger(\tau) \vec{z}_{(0)}=  \exp \left(\frac{2 \pi \im x_*}{NL} f(\tau)\right) \vec{z}_{(0)}, 
\end{align}
where $f(\tau)$ is any smooth function with
\begin{align}
    f(\tau) \to 
    \begin{cases}
    0 & \quad \text{as} \quad \tau \to - \infty, \\
    1 & \quad \text{as} \quad \tau \to + \infty. 
    \end{cases}
\end{align}
On making the gauge transformation $z \mapsto h^\dagger z$, the theta term is transformed as (see Appendix~\ref{eq:theta_cylinder})
\begin{align}
    \frac{1}{2\pi} \int `` \theta\, \diff a[\vec{z}_{(0)}] "
    \mapsto
    \frac{1}{2\pi} \int `` \theta\, \diff a[\vec{z}_{(0)}] "
    - \frac{2 \pi w x_*}{NL},  
\end{align}
where $w$ is the winding number, $\theta(L)-\theta(0)=2\pi w$. 
We thus find that the imaginary part of the fractional instanton action depends on $x_*$ through the term $ 2 \pi w x_*/NL    \eqcolon  w \varphi_*$, which is enough to see the vanishing of the imaginary-time transition amplitude $\langle 1 | \exp(-\beta \hat{H})|0 \rangle$: 
\begin{align}
\langle {1}| {\rm e}^{-\beta \hat{H}} | 0\rangle  = 
  {\rm e}^{-\frac{S_I}{N} + {\rm i} \frac {\bar \theta}{N} } \int _0^{2 \pi} {\rm d} \varphi_* \;    {\rm e}^{ {\rm i} w \varphi_*}=0 ,
\end{align}
for $w\not=0$. 
This means fractional instantons do not lift the degeneracy of the states $|k\rangle$ at the leading order of semiclassics. 
As a result, we obtain $N$-fold degeneracy from the semiclassical analysis, which is the minimal degeneracy required to match the anomaly when the winding of $\theta$ satisfies $\mathrm{gcd}(N,w)=1$.


\subsection{Flavor charges due to the winding \texorpdfstring{$\theta$}{theta} and exact \texorpdfstring{$N$}{N}-fold degeneracy}
\label{sec:semiclassics_windingtheta}

In Sec.~\ref{sec:destructive_interference}, we have observed the $N$-fold degeneracy of ground states at the leading order of semiclassics whenever $w\not=0$. 
This is contrary to the minimal requirement for the anomaly matching because it only requires the $\frac{N}{\mathrm{gcd}(N,w)}$ degeneracy as discussed in Sec.~\ref{sec:FTBC_Anomalies}.  
Since 
\begin{align}
    N=\mathrm{gcd}(N,w)\cdot \frac{N}{\mathrm{gcd}(N,w)}, 
\end{align} 
the $N$-fold degeneracy is perfectly consistent with the anomaly. Nevertheless, it is still somewhat surprising that we should have a degeneracy of ground states that is not the minimal one required. 
For $\mathrm{gcd}(N,w)>1$, it is tempting to suspect that the classical degeneracy is partially lifted at a higher order in semiclassics, as would be detected by the nonvanishing of some transition amplitudes between nonconsecutive classical vacua.

However, this does not happen -- the vanishing of all transition amplitudes $\langle k' | \rme^{-\beta \hat{H}}|k\rangle$ turns out to be exact. As we will now show, in the presence of a winding $\theta$ angle, the states $|k \rangle$ acquire distinct charges under the $U(1)^{N-1}$ global symmetry, and hence all transitions between them are forbidden.

Let us consider the charge $Q_T$ associated with the $U(1)$ symmetry $\vec{z} \mapsto \rme^{\im \alpha T} \vec{z}$, where $T$ is a diagonal matrix.  Using the standard Noether formula
\begin{align}
    Q_T = - \int \bigg\{ 
    \frac{\delta L}{\delta (\partial_t \vec{z}\,)} \delta \vec{z}
    + \delta \vec{z}^{\, \dagger} \frac{\delta L}{\delta (\partial_t \vec{z}^{\, \dagger})}
    \bigg\} \, \diff x
\end{align}
with $\delta \vec{z} = \im T \vec{z}$, we find
\begin{align}
    Q_T = - \frac{1}{g^2} \int \bigg\{
    \im (\partial_t \vec{z}^{\, \dagger}T \vec{z} - \vec{z}^{\, \dagger} T \partial_t \vec{z}\,) + 2 a_t (\vec{z}^{\, \dagger} T \vec{z} \,) \bigg\} \, \diff x.
\end{align}
Here, $a_t$ is determined by its classical equation of motion, which is modified by the nontrivial spatial dependence of $\theta$: 
\begin{align}
    a_t = \im \vec{z}^{\,\dagger} \partial_t \vec{z} - \frac{g^2}{4 \pi} \partial_x \theta. 
\end{align}
The Noether charge can thus be rewritten
\begin{align}
    Q_T = - \frac{\im}{g^2} \int \bigg\{
    \partial_t \vec{z}^{\, \dagger}T \vec{z} - \vec{z}^{\, \dagger} T \partial_t \vec{z} + 2 (\vec{z}^{\,\dagger} \partial_t \vec{z}\,) (\vec{z}^{\, \dagger} T \vec{z}\,) \bigg\} \, \diff x
    + \frac{1}{2 \pi} \int \partial_x \theta\, (\vec{z}^{\, \dagger} T \vec{z}\,) \, \diff x. 
\end{align}
Substituting $\vec{z} = \vec{\zeta}_{(k)}$, we find the charge of the state $|k\rangle$ to be given by
\begin{align}
    Q_T[\zeta_{(k)}] = w \left( \vec{\zeta}_{(k)}^{\,\dagger} T \vec{\zeta}_{(k)}^{\vphantom{\dagger}} \right).
\end{align}
This establishes that the classical vacua $\vec{\zeta}_{(k)}$ acquire distinct charges under the $U(1)$ symmetry generated by $T$ due to the  winding of the $\theta$ angle.
In particular, choosing $T = \mathrm{diag}(0,1,\ldots,N-1)$, we have ($Q  \coloneq Q_{\mathrm{diag}(0,1,\ldots, N-1)}$) 
\begin{align}
Q[\vec{\zeta}_{(k)}] = wk.
\label{eq:statecharging}
\end{align}
As the states $|k\rangle$ (with $k=0,1,\ldots, N-1$) have different $U(1)^{N-1}$ charges, the tunneling processes between them do not occur at all.

To get a better understanding, let us discuss a more elementary example showing the same phenomenon: the single $SU(2)$ spin system with the Hamiltonian
\begin{align}
    \hat{H}=J\hat{S}_z^2, 
    \label{eq:spin_Hamiltonian}
\end{align}
which breaks the $SU(2)/\mathbb{Z}_2\simeq SO(3)$ spin rotational symmetry to $(U(1)/\mathbb{Z}_2)\rtimes \mathbb{Z}_2\simeq O(2)$. 
For half-integer spins, the $O(2)$ symmetry has an 't~Hooft anomaly, while for integer spins there is no anomaly. 
The crucial difference between them is whether the $S_z=0$ state exists or not, which gives a singlet state. 
When $J>0$ (the so-called easy-plane case), it strongly affects the ground-state structure: For half-integer spins, the ground states are two-fold degenerate given by $S_z=\pm \frac{1}{2}$, while for integer spins, we have the unique ground state, $S_z=0$. 
This realizes the minimal requirement of the 't~Hooft anomaly of the $O(2)$ symmetry.
When $J<0$ (the so-called easy-axis case), however, we always have two-fold degeneracy given by $S_z=\pm s$ whether the spin $s$ is integer or half-integer. 
Thus, the easy-axis case is similar to our current situation in the $\mathbb{C}P^{N-1}$ model on $\mathbb{R}\times S^1$ with the flavor-twisted boundary condition and a winding $\theta$ angle. 

We can see the similarity more clearly by looking at the path integral for \eqref{eq:spin_Hamiltonian} with spin $s$. The action is given by 
\begin{align}
    S[\vartheta,\phi]=\im\,(2 s) \int \sin^2\frac{\vartheta}{2}\, \diff \phi+\int \diff \tau \left(\frac{m}{2}(\dot{\vartheta}^2+\sin^2\vartheta\,\, \dot{\phi}^2)+J s^2 \cos^2 \vartheta\right). 
\end{align}
Here, we take the polar coordinate, $\vec{z}=(\cos \frac{\vartheta}{2}, \rme^{\im \phi}\sin \frac{\vartheta}{2})$, and $m=s\,\delta \tau/2\,(\,\to 0)$. 
The $U(1)/\mathbb{Z}_2$ symmetry shifts $\phi\mapsto \phi+\alpha$, and the $\mathbb{Z}_2$ symmetry acts as the charge conjugation, $\phi\mapsto-\phi$ and $\vartheta\mapsto \pi-\vartheta$. 
The classical vacua for $J<0$ are given by $\vartheta=0,\pi$, which correspond to $\vec{z}_{(0)}\sim (1,0)$ and $\vec{z}_{(1)}\sim (0,1)$, respectively. 
Whenever $s\not =0$, the destructive interference of the Berry phase gives the complete annihilation of the transition amplitude due to the continuous modulus $\phi$ associated with the $U(1)$ spin symmetry, and it occurs exactly in the same way as we discussed in Sec.~\ref{sec:destructive_interference}. 
This is the semiclassical manifestation of the fact that the Berry phase assigns the different $U(1)$ charges to the classical vacua, and thus those states, $\vec{z}_{(0)}$ and $\vec{z}_{(1)}$, cannot be mixed unless the $U(1)$ symmetry is broken to a smaller subgroup.

Upon adding some perturbations that break the $PSU(N)$ symmetry down to $\mathbb{Z}_N\times \mathbb{Z}_N$, we should get the minimal degeneracy $\frac{N}{\mathrm{gcd}(N,w)}$ required by the anomaly. 
Indeed, even when $Q$ itself is no longer a conserved charge, \eqref{eq:statecharging} still implies that the state $|k\rangle$ has the eigenvalue $\rme^{\frac{2\pi \im }{N} wk}$ for the clock symmetry operator, and the selection rule is relaxed to allow transitions $|k\rangle \to |k'\rangle$ with $wk = wk' \mod N$. Similarly, in the above $SU(2)$ spin example, we can resolve the degeneracy for the integer spin by a small perturbation, $J_x \hat{S}_x^2\sim J_x s^2 \sin^2\vartheta \cos(2\phi)$, but this perturbation does not lift the degeneracy for the half-integer spins as we have seen in Sec.~\ref{sec:CPN_QM_Berry}. 

In summary, the degeneracy of $N$ we have obtained is exact, and independent of the profile and winding number $w$ of the $\theta$ parameter in the sufficiently small circle regime. For the case of $w=1$ and a monotonic profile of $\theta$, it seems plausible that this degeneracy is maintained as the circle size is increased and the volume independence may work. 
However, we cannot say much about what happens for the case of generic winding numbers $w$ and generic profiles of $\theta$ and they would require more detailed analysis on each case.

\section{Summary and discussions}

We have discussed the $\theta$ interface of the $2$d $\mathbb{C}P^{N-1}$ sigma model from various perspectives. 
If the spatial direction is sufficiently large, then the ground-state wave function is trivially gapped inside the bulk, but low-energy excitations still exist and localize on the wall. 
We provide its intuitive understanding by realizing the $\mathbb{C}P^{N-1}$ sigma model as the effective theory of the $SU(N)$ spin chain. 
The two-fold degeneracy at $\theta=\pi$ is then associated with the spontaneous breaking of lattice translation, and the $\theta$ interface should have a free $SU(N)$ spin. 
As a model of its dynamics, we consider the $\mathbb{C}P^{N-1}$ quantum mechanics with the Berry phase and have studied its properties to confirm the $N$ degenerate vacua. 

We also considered the small $S^1$ compactification with the flavor-twisted boundary condition and nontrivially winding $\theta$ angle. 
Due to the flavor twist, we can use reliable semiclassical computations to study the asymptotically free field theory. 
At the classical level, there are $N$ degenerate vacua whether or not $\theta$ has nonzero winding number. 
When there is no winding of $\theta$, the dilute gas approximation of fractional instantons shows the uniqueness of the ground state at generic values of $\theta$ so that we also get the $N$-branch structure of the vacua as we expect from the adiabatic continuity. 
We have confirmed that this story becomes totally different when $\theta$ has nonzero winding number. 
In such cases, the moduli integral of the fractional instanton gives the complete cancellation of the transition amplitude, and the $N$-fold degeneracy remains even after taking into account the effects of fractional instantons. 

Although we have focused on the $2$d $\mathbb{C}P^{N-1}$ sigma model in this paper, our observation itself should be quite general and we believe that our semiclassical techniques can be extended to many other cases. 
As a straightforward extension, we can consider the $2$d $SU(N)/U(1)^{N-1}$ sigma model instead of $\mathbb{C}P^{N-1}$. 
This theory has $N-1$ independent topological charges, so we have many $\theta$ parameters, $\theta_{i=1,\ldots, N-1}$, to be discussed. 
We can obtain this theory as the low-energy effective theory of an anti-ferromagnetic $SU(N)$ spin chain~\cite{Bykov:2011ai, Bykov:2012am, Lajko:2017wif}. 
As it has an 't~Hooft anomaly between $PSU(N)$ and the $\theta$ periodicity~\cite{Tanizaki:2018xto}, the $\theta$ interface should support a projective representation so we must obtain the $N$-fold degeneracy. 
When we perform the $S^1$ compactification with the flavor-twisted boundary condition, we have $N!$ classical vacua instead of $N$, and the transition amplitudes can be computed by using fractional instantons characterized by $N-1$ topological charges~\cite{Hongo:2018rpy}. 
It would be interesting to see if we can obtain the $N$ degenerate vacua out of $N!$ classical vacua when one of the $\theta$s has a winding number. 

A more challenging subject is to extend this work to the case of $4$d gauge theories with a winding $\theta$ angle. 
Some recent works try to achieve the adiabatic continuity of various compactified setups of $4$d gauge theories~\cite{Yamazaki:2017ulc, Cox:2021vsa,  Tanizaki:2022ngt, Tanizaki:2022plm}. 
For example, Ref.~\cite{Yamazaki:2017ulc} shows the connection between the $4$d Yang-Mills theory on $\mathbb{R}\times T^3$ with the 't~Hooft flux and the $2$d $\mathbb{C}P^{N-1}$ sigma model on $\mathbb{R}\times S^1$ with the flavor-twisted boundary condition and Ref.~\cite{Tanizaki:2022ngt} relates the $4$d massless quantum chromodynamics (QCD) and the $2$d Wess-Zumino-Witten model. 
These observations strongly suggest that we can study the physics of the $\theta$ interface of $4$d gauge theories also by a suitable compactification with winding $\theta$ using the reliable semiclassics.

We suspect that the $S^1$ compactification with winding $\theta$ has an application in the semiclassical study of $4$d chiral gauge theories on $\mathbb{R}^3\times S^1$. 
Applying the double-trace deformation to the gauge sector, we can study the confinement dynamics of $4$d gauge theories by the dilute gas of monopole-instantons~\cite{Unsal:2007vu, Unsal:2008ch, Shifman:2008ja, Poppitz:2012sw, Poppitz:2012nz}, but it always requires the gapping out of the dynamical electric charges by a suitable choice of boundary condition. 
In the case of chiral gauge theories, we have to introduce the chiral-symmetric mass by using the twisted boundary condition of chiral symmetry. 
In such cases, due to the Adler-Bell-Jackiw (ABJ) anomaly, it should be related to the situation with a nonzero winding $\theta$ angle. 
As we have seen, such a winding $\theta$ angle can drastically change the consequence of the semiclassical analysis. 
Since this possibility has been overlooked in previous studies~\cite{Shifman:2008cx, Sulejmanpasic:2020zfs}, it would be interesting to reconsider the chiral gauge theories on $\mathbb{R}^3\times S^1$ taking into account the effect of winding $\theta$.  

Even in the study of vector-like theories, the QCD-like theories with higher representation fermions possess discrete chiral symmetries.  In such cases, one can consider chirally twisted boundary conditions to study the physics on the interface. As emphasized above, due to the ABJ anomaly, this can also be formulated as a winding theta when the theory is formulated on $\mathbb R^3 \times S^1$.   
If we consider these theories in set-ups where the center-symmetry acting on the Polyakov loop is stable and the theory abelianizes at long distances,  one lands on theories with monopole-instantons in the presence of Chern-Simons terms~\cite{Poppitz:2008hr, Poppitz:2020tto} (The case with winding $\theta$ and broken center symmetry is recently studied in  \cite{Kan:2019rsz}). 
These regimes  should lead to deconfinement despite the presence of monopoles~\cite{Pisarski:1986gr, Affleck:1989qf, Lee:1991ge}. It would be interesting to understand the microscopic mechanism through which this takes place.

\acknowledgments

The authors thank Hanqing~Liu, Tatsuhiro~Misumi, Erich~Poppitz, Hersh~Singh and Tin~Sulejmanpasic for useful discussions. 
The work of Y.~T. was supported by Japan Society for the Promotion of Science (JSPS) KAKENHI Grant numbers, 22H01218 and 20K22350, and by Center for Gravitational Physics and Quantum Information (CGPQI) at Yukawa Institute for Theoretical Physics.
The work of M.~\"{U}. was supported by U.S. Department of Energy, Office of Science, Office of Nuclear Physics under Award Number DE-FG02-03ER41260.

\appendix

\section{Spacetime-dependent \texorpdfstring{$\theta$}{theta} term of the \texorpdfstring{$2$}{2}d \texorpdfstring{$\mathbb{C}P^{N-1}$}{CP(N-1)} model}
\label{sec:spcetimedependent_theta}

In this appendix, following Ref.~\cite{Cordova:2019jnf}, we discuss the definition of the $\theta$ term,
\begin{equation}
    \frac{\im}{2\pi}\int_{M_2} \theta\, \diff a,
\end{equation}
for the $2$d $\mathbb{C}P^{N-1}$ model when the $\theta$ parameter has the spacetime dependence, i.e. $\theta=\theta(x)$ is no longer a constant in terms of $x\in M_2$. 
We would like to maintain the identification, $\theta\sim \theta+2\pi$, so that $\theta(x)$ can be regarded as a background $2\pi$-periodic scalar field. 
Then, the $2\pi$ periodicity, $\theta\sim \theta+2\pi$, is a kind of the gauge redundancy, so the integrand $\theta \diff a$ is not gauge invariant. We shall give a gauge-invariant definition on general $2$d oriented closed manifolds $M_2$ by using its simplicial decomposition. 
Based on the gauge-invariant definition, we also describe the explicit form for the case, where $M_2=S^1 \times \mathbb{R}$ and $\theta$ has the nontrivial winding along the circle. 

\subsection{Definition of the spacetime-dependent \texorpdfstring{$\theta$}{theta} term}

As we have seen above, the problem of the spacetime-dependent $\theta$ term is very similar to the problem of defining the $3$d Chern-Simons action, $\mathrm{CS}[a]=``\frac{\im}{4\pi}\int_{M_3} a\diff a"$, where we put the quotation mark to remember that the right-hand-side is the heuristic definition. 
We usually resolve this subtlety by regarding $M_3$ as the boundary of $M_4$, $\partial M_4=M_3$, and set 
\begin{equation}
    \mathrm{CS}[a]=\frac{\im}{4\pi}\int_{M_4} \diff \tilde{a} \wedge \diff \tilde{a}, 
\end{equation}
where $\tilde{a}$ is a $U(1)$ gauge field on $M_4$ with $\tilde{a}|_{\partial M_4}=a$.
This is manifestly gauge invariant, and it is independent of the choice of spin $4$-manifolds $M_4$ and extensions $\tilde{a}$ in mod $2\pi \im$, which is sufficient to define the path-integral weight. 
Therefore, one may think that we can apply the same trick, and claim that the following definition should work,
\begin{equation}
    \frac{\im}{2\pi}\int_{M_2} ``\,\theta\, \diff a\,"\overset{?}{=} \frac{\im}{2\pi}\int_{M_3} \diff \tilde{\theta} \wedge \diff \tilde{a},
    \label{eq:trial_def}
\end{equation}
where $M_3$ is a $3$d manifold with $\partial M_3=M_2$, and $\tilde{\theta}$ and $\tilde{a}$ are the extensions of $\theta$ and $a$ to $M_3$, respectively. 
However, this does not work because of the following reasons. 

In the case of $3$d Chern-Simons action, $3$d $U(1)$ gauge fields always have an extension to a $4$d manifold, which is ensured by the bordism group, $\Omega_3^{\mathrm{spin}}(\mathcal{B}U(1))=0$.  
This is why we can define the $3$d Chern-Simons action by extending the spacetime manifolds. 
On the other hand, $2$d $U(1)$ gauge fields, or $2$d $\mathbb{C}P^{N-1}$ fields, do not necessarily have the $3$d extension as $\Omega^{\mathrm{spin}}_2(\mathbb{C}P^{N-1})\simeq \Omega^{\mathrm{spin}}_{2}(\calB U(1))\simeq \mathbb{Z}$. 
To see this obstruction explicitly, let us consider the $U(1)$ gauge field on $S^2\subset \mathbb{R}^3$ with the Dirac monopole at the origin of $\mathbb{R}^3$. 
Due to the monopole singularity at the origin, we cannot have its $3$d extension as a smooth $U(1)$ gauge field. 
As a result, we cannot use \eqref{eq:trial_def} when the $U(1)$ gauge field has a nontrivial instanton number, and we need to discuss another way to define the spacetime dependent $\theta$ term.

Let us approximate the closed $2$-manifold $M_2$ by a polyhedron.
That is, $\sigma_i$ is a polygon including its boundary, $\sigma_{ij}=\sigma_i\cap \sigma_j$ is a connected line segment (or empty), and $\sigma_{ijk}=\sigma_{ij}\cap \sigma_{jk}\cap \sigma_{ki}$ is a point (or empty), and $\bigcup_i \sigma_i$ is homeomorphic to $M_2$. 
Since we are going to define the topological term $\frac{\im}{2\pi}\int_{M_2} ``\theta\, \diff a"$, we just identify $M_2=\bigcup_i \sigma_i$ for simplicity of notation. 
Let us assume that $M_2$ is oriented, then the orientation of $2$-simplex $\sigma_i$ is chosen consistently. For $i<j$, the orientation of $\sigma_{ij}(\subset \partial \sigma_i)$ is fixed by $\sigma_i$ in a canonical way, and $\sigma_{ji}$ has the opposite orientation. 
In the following explanation, we assume that there is no quadruple overlap, but the extension to such cases should be straightforward.

Next, we describe the $2\pi$-periodic scalar $\theta$ and the $U(1)$ gauge field on each patch of the polyhedron $\bigcup_i \sigma_i$. 
The $\theta$ field consists of the following data,
\begin{itemize}
    \item $\theta_i:\sigma_i\to \mathbb{R}$, 
    \item $w_{ij}: \sigma_{ij}\to \mathbb{Z}$,
\end{itemize}
which satisfy 
\begin{equation}
    \left.(\theta_i-\theta_j)\right|_{\sigma_{ij}}=2\pi w_{ij}. 
\end{equation}
The consistency requires that $w_{ji}=-w_{ij}$ and 
\begin{equation}
    w_{ij}+w_{jk}+w_{ki}=\left.\frac{1}{2\pi}\left((\theta_i-\theta_j)+(\theta_j-\theta_k)+(\theta_k-\theta_i)\right)\right|_{\sigma_{ijk}}=0. 
\end{equation}
Its $2\pi$ periodicity is realized by postulating the gauge invariance under 
\begin{equation}
    \theta_i\mapsto \theta_i+2\pi \omega_i, \quad w_{ij}\mapsto w_{ij}+\omega_i-\omega_j, 
\end{equation}
where $\omega_i\in \mathbb{Z}$ is the gauge parameter. We denote $(\delta \omega)_{ij}=\omega_i-\omega_j$, then $w_{ij}\mapsto w_{ij}+(\delta \omega)_{ij}$. 

The $U(1)$ gauge field $a$ consists of the data
\begin{itemize}
    \item $a_i$: an $\mathbb{R}$-valued $1$-form on $\sigma_i$,
    \item $g_{ij}:\sigma_{ij}\to U(1)$, 
\end{itemize}
which satisfy 
\begin{equation}
    a_j=a_i-\im g_{ij}^{-1}\diff g_{ij}
\end{equation}
on $\sigma_{ij}$. On $\sigma_{ijk}$, we impose the cocycle condition, 
\begin{equation}
    g_{ij}g_{jk}g_{ki}=1. 
\end{equation}
Although this is one of the standard ways to define the $U(1)$ gauge field, it turns out that this does not give enough data to define $``\theta \diff a"$. 
We should take an $\mathbb{R}$-valued lift of the transition function $g_{ij}$, so we need the following data 
\begin{itemize}
    \item $\phi_{ij}:\sigma_{ij}\to \mathbb{R}$, 
    \item $n_{ijk}:\sigma_{ijk}\to \mathbb{Z}$, 
\end{itemize}
that satisfy $g_{ij}=\rme^{-\im \phi_{ij}}$ and 
\begin{equation}
    \phi_{ij}+\phi_{jk}+\phi_{ki}=2\pi n_{ijk}
\end{equation}
on $\sigma_{ijk}$. We can rewrite the connection formula for $\{a_i\}_i$ as $a_i-a_j=\diff \phi_{ij}$. 
The gauge identification is given by 
\begin{equation}
    a_i\mapsto a_i+\diff \alpha_i,\quad \phi_{ij}\mapsto \phi_{ij}+(\delta \alpha)_{ij}+2\pi \nu_{ij}, \quad 
    n_{ijk}\mapsto n_{ijk}+(\delta \nu)_{ijk},
\end{equation}
where $\alpha_i:\sigma_i\to \mathbb{R}$, $\nu_{ij}\in \mathbb{Z}$ and $(\delta \nu)_{ijk}=\nu_{ij}-\nu_{ik}+\nu_{jk}$. 
As an example, we can compute the topological charge in the following way, 
\begin{align}
    \frac{1}{2\pi}\int_{M_2}\diff a
    &= \sum_{i}\frac{1}{2\pi}\int_{\sigma_i}\diff a_i = \sum_i \frac{1}{2\pi}\int_{\partial \sigma_i}a_i\notag\\
    &= \sum_{i<j} \frac{1}{2\pi}\int_{\sigma_{ij}} (\delta a)_{ij}=\sum_{i<j}\frac{1}{2\pi}\int_{\sigma_{ij}}\diff \phi_{ij} \notag\\
    &=\sum_{i<j} \frac{1}{2\pi}\int_{\partial \sigma_{ij}} \phi_{ij} =\sum_{i<j<k}\frac{1}{2\pi}\int_{\mathrm{sign}(\sigma_{ijk})\sigma_{ijk}} (\delta \phi)_{ijk}\notag\\
    &=\sum_{i<j<k} \mathrm{sign}(\sigma_{ijk}) n_{ijk},
\end{align}
where $\mathrm{sign}(\sigma_{ijk})$ is the orientation of $\sigma_{ijk}$ determined from $\partial \sigma_{ij}$ for $i<j<k$. 

The gauge-invariant definition of the spacetime-dependent $\theta$ term is given by~\cite{Cordova:2019jnf} 
\begin{align}
    \frac{1}{2\pi}\int_{M_2}``\theta\,\diff a" = \sum_{i}\frac{1}{2\pi}\int_{\sigma_i}\theta_i \diff a_i-\sum_{i<j}\int_{\sigma_{ij}}w_{ij}a_j +\sum_{i<j<k}\int_{\sign(\sigma_{ijk})\sigma_{ijk}}w_{ij}\phi_{jk}. 
    \label{eq:def_theta}
\end{align}
For constant $\theta$, this reduces to the original definition as we can set $\theta_i=\theta$ and $w_{ij}=0$. 
Let us check the gauge invariance of this expression. The first term on the right-hand-side transforms as 
\begin{align}
    &\quad \sum_i \frac{1}{2\pi}\int_{\sigma_i}(\theta_i+2\pi \omega_i) \diff (a_i+\diff \alpha_i)-\sum_i \frac{1}{2\pi}\int_{\sigma_i} \theta_i \diff a_i \notag\\
    &=\sum_i \omega_i \int_{\partial \sigma_i} a_i \notag\\
    &=\sum_{i<j} \int_{\sigma_{ij}} (\omega_i a_i-\omega_j a_j) \notag\\
    &=\sum_{i<j}\left( \int_{\sigma_{ij}}(\omega_i-\omega_j) a_j + \int_{\partial \sigma_{ij}} \omega_i \phi_{ij}\right) \notag\\
    &=\sum_{i<j}\int_{\sigma_{ij}}(\omega_i-\omega_j)a_j+\sum_{i<j<k}\int_{\sign(\sigma_{ijk})\sigma_{ijk}}(\omega_i\phi_{ij}-\omega_i \phi_{ik}+\omega_j \phi_{jk})\notag\\
    &=\sum_{i<j}\int_{\sigma_{ij}}(\omega_i-\omega_j)a_j-\sum_{i<j<k}\int_{\sign(\sigma_{ijk})\sigma_{ijk}} (\omega_i-\omega_j)\phi_{jk}\quad  (\bmod\,\, 2\pi \mathbb{Z}). 
    \label{eq:gauge_variation}
\end{align}
In the last step, we used $(\delta \phi)_{ijk}=2\pi n_{ijk}$. 
Now the necessity of the second and third term on the right-hand-side of \eqref{eq:def_theta} becomes evident by noting that $w_{ij}$ transforms as $w_{ij}\mapsto w_{ij}+(\omega_i-\omega_j)$: The second term of \eqref{eq:def_theta} is necessary to cancel the first term of \eqref{eq:gauge_variation}, and the third term of \eqref{eq:def_theta} is also necessary to cancel the last term of \eqref{eq:gauge_variation}. 
To complete the discussion, we also need to check if the second and third terms of \eqref{eq:def_theta} is invariant under the $U(1)$ gauge transformation. 
The second term of \eqref{eq:def_theta} transforms as 
\begin{align}
    &\quad -\sum_{i<j}\int_{\sigma_{ij}} (w_{ij} (a_j+\diff \alpha_j)- w_{ij } a_j)\notag\\
    &=-\sum_{i<j}\int_{\partial \sigma_{ij}} w_{ij}\alpha_j \notag\\
    &=-\sum_{i<j<k}\int_{\sign(\sigma_{ijk}) \sigma_{ijk}} (w_{ij}\alpha_j-w_{ik}\alpha_{k}+w_{jk}\alpha_k) \notag\\
    &=-\sum_{i<j<k}\int_{\sign(\sigma_{ijk})\sigma_{ijk}} w_{ij} (\alpha_i-\alpha_j). 
\end{align}
As $\phi_{ij}$ transforms as $\phi_{ij}\mapsto \phi_{ij}+(\delta \alpha)_{ij}+2\pi \nu_{ij}$, this is exactly cancelled by the gauge variation of the third term of \eqref{eq:def_theta}. 
We have confirmed that the definition \eqref{eq:def_theta} is gauge invariant mod $2\pi$, and we can use it to define the path-integral weight.

\subsection{Concrete expression for the cylinder}
\label{eq:theta_cylinder}

Consider the cylinder $S^1 \times \mathbb{R}$ with circumference $L$. If we regard $\theta$ as an $S^1$-valued function on $[0,L] \times \mathbb{R}$ such that $\theta(L,\tau) = \theta(0,\tau)$, then we can lift it to a real-valued function $\hat{\theta}$ on $[0,L] \times \mathbb{R}$ such that $\hat{\theta}(L,\tau) = \hat{\theta}(0,\tau) + 2 \pi w$, where $w$ is the winding number of $\theta$. Note also that on a cylinder, it is always possible to choose the transition functions of the $U(1)$ gauge field $a$ to be trivial: $g_{ij} = 1$. That is to say, it is possible to choose $a_i = \hat{a}|_{\sigma_i}$ for some global 1-form $\hat{a}$. We shall show that the theta term \eqref{eq:def_theta} can be rewritten just in terms of $\hat{\theta}$ and $\hat{a}$. 

A good cover of the cylinder can be achieved by taking three rectangular cells $\sigma_{1,2,3}$, say
\begin{equation}
    \sigma_1 =[0,L/3] \times \mathbb{R}, \quad \sigma_2 = [L/3,2L/3] \times  \mathbb{R}, \quad \sigma_3 =  [2L/3,L] \times \mathbb{R},
\end{equation}
With respect to this cover, there are just three transition functions for $\theta$: $w_{12},w_{23},w_{31}$. However, if we choose $\theta_i = \hat{\theta}|_{\sigma_i}$, then $w_{12}$ and $w_{23}$ vanish, and $w_{31} = w$. With this choice, the theta term \eqref{eq:def_theta} reduces to the following simple form:
\begin{equation}
    \frac{1}{2\pi} \int_{S^1 \times \mathbb{R}} `` \theta\, \diff a "
    = \frac{1}{2 \pi} \int_{[0,L] \times \mathbb{R}} \hat{\theta} \, \diff \hat{a}
    - w \int_{-\infty}^{+\infty} \hat{a}_\tau(0,\tau) \, \diff \tau.
    \label{eq:def_theta_cylinder}
\end{equation}

It is very important to realize that the theta term on the cylinder is not fully gauge invariant: under a gauge transformation $a \mapsto a + \diff \alpha$, we have
\begin{align}
    \frac{1}{2\pi} \int_{S^1 \times \mathbb{R}} `` \theta\, \diff a "
    \mapsto
    \frac{1}{2\pi} \int_{S^1 \times \mathbb{R}} `` \theta\, \diff a "
    -w\{\alpha(0,+ \infty) - \alpha(0,-\infty)\}.
    \label{eq:ThetaGaugeTransform}
\end{align}
Thus, the theta term can gauge transform nontrivially only if 
(1) the winding number $w$ is $\neq 0$, and (2) $\alpha(0,+ \infty) - \alpha(0,-\infty) \neq 0$, i.e. the gauge transformation is ``large'' in time. This simple observation will be crucial for the present work.

\section{\texorpdfstring{$SU(N)$}{SU(N)} spin chain and \texorpdfstring{$2$d $\mathbb{C}P^{N-1}$}{2d CP{N-1}} sigma model}
\label{sec:CPsigma_SUNchain}

Here, we give a review about the realization of the $2$d $\mathbb{C}P^{N-1}$ sigma model around $\theta=\pi$ by the $SU(N)$ spin chain. 
For $N=2$, it is found by Haldane~\cite{Haldane:1982rj,Haldane:1983ru} that the $SU(2)$ antiferromagnetic spin chain of spin $S$ gives the $\mathbb{C}P^1$ sigma model with $\theta=2\pi S$. 
Its extension to the $2$d $\mathbb{C}P^{N-1}$ model has been discussed in the context of $D$-theory~\cite{Brower:2003vy, Beard:2004jr, Caspar:2022llo}. 

\subsection{Spin coherent state and path integral}
\label{sec:BerryPhase}

We denote the $SU(N)$ spin operator as $\hat{S}_{\alpha \beta}$ with $\alpha,\beta=1,\ldots, N$. For $U\in SU(N)$, they transform as $\hat{S}_{\alpha \beta}\to (U\hat{S} U^\dagger)_{\alpha \beta}$. 
They satisfy $\hat{S}^\dagger_{\alpha \beta}=\hat{S}_{\beta \alpha}$ and the commutation relation, 
\begin{equation}
    [\hat{S}_{\alpha_1 \alpha_2}, \hat{S}_{\beta_1 \beta_2}]=\delta_{\alpha_2 \beta_1}\hat{S}_{\alpha_1 \beta_2}-\delta_{\beta_2\alpha_1}\hat{S}_{\beta_1 \alpha_2}. 
\end{equation}
We introduce the Schwinger boson to conveniently describe the spin coherent state~\cite{Arovas:1988zza}. The Schwinger boson is the $N$-component harmonic oscillator, whose creation and annihilation operators are $\hat{a}^\dagger_{\alpha}$ and $\hat{a}_\alpha$, respectively, with the commutation relation, 
\begin{equation}
    [\hat{a}_{\alpha},\hat{a}^\dagger_\beta]=\delta_{\alpha \beta}. 
\end{equation}
We can readily check that 
\begin{equation}
    \hat{S}_{\alpha \beta}=\hat{a}^\dagger_\alpha \hat{a}_\beta
\end{equation}
satisfies the above $SU(N)$ commutation relation. 
In this convention, $\hat{a}^\dagger_\alpha$ transforms as the defining representation of $SU(N)$ and $\hat{a}_\alpha$ transforms as its conjugate representation.

Let $|0\rangle$ be the vacuum of the harmonic oscillators annihilated by $\hat{a}_{\alpha}$ for $\alpha=1,\ldots, N$. This belongs to the trivial representation of $SU(N)$. 
Applying $p$ creation operators to the vacuum, we find that those states,
\begin{align}
    \hat{a}^\dagger_{\alpha_1}\cdots \hat{a}^\dagger_{\alpha_p}|0\rangle,
\end{align}
form the totally symmetric $p$-box representation of $SU(N)$. 
Let $\vec{\phi}\in \mathbb{C}^N$ with $|\vec{\phi}\,|^2=1$, then we define the coherent state for the $p$-box symmetric representation by 
\begin{align}
    |\vec{\phi}\,\rangle=\frac{1}{\sqrt{p!}}(\phi_1 \hat{a}^\dagger_1+\cdots +\phi_N \hat{a}^\dagger_N)^p |0\rangle. 
\end{align}
It satisfies the following property, 
\begin{align}
    &\langle \vec{\phi}'\,| \vec{\phi}\,\rangle = (\vec{\phi}'^\dagger\cdot \vec{\phi}\,)^p,\\
    &\langle \vec{\phi}\,| \hat{S}_{\alpha\beta}| \vec{\phi}\,\rangle = p\, \phi^*_\alpha \phi_\beta,\\
    &\int \diff \Omega_\phi |\vec{\phi}\,\rangle \langle \vec{\phi}\,|=\bm{1}. 
\end{align}
Using these formula, we can, for example, obtain the path integral formula for the Hamiltonian $\hat{H}=J_{\alpha \beta}\hat{S}_{\alpha \beta}$ as follows:
\begin{align}
    \mathrm{Tr}(\exp(-\beta\hat{H}))=\int \Diff \phi \exp(-S[\phi]),
\end{align}
where the Euclidean action is given by
\begin{align}
    S[\phi]=\int_{0}^\beta p \,\vec{\phi}\,^\dagger \cdot \diff \vec{\phi}+\int_{0}^\beta p\,J_{\alpha \beta}\phi^*_\alpha \phi_\beta \diff \tau. 
\end{align}
The first term is the Berry phase, and the second term describes the classical energy. 
Strictly speaking, we use the semiclassical approximation to describe the path integral expression, and this approximation is valid in the limit $p\to \infty$. 

\subsection{Low-energy limit of spin chains}

To obtain the $\mathbb{C}P^{N-1}$ sigma model, we define the spin chain with even number of sites, $i=1,2,\ldots, 2L$, and we put the $p$-box symmetric representation on the odd sites, $i=1,3,\ldots, 2L-1$ and put its conjugate representation on the even sites, $i=2,4,\ldots, 2L$.

Let us denote the spinon field on the odd site $2n-1$ as $\vec{\phi}^{\,1}_n$ and the one on the odd site $2n$ as $\vec{\phi}^{\,2}_n$, then the Euclidean action becomes 
\begin{align}
    S&=\int_0^\beta \diff \tau\sum_{n=1}^{L}\left\{
    p\, (\vec{\phi}^{\,1\dagger}_n\cdot \partial_\tau \vec{\phi}^{\,1}_n-\vec{\phi}^{\,2\dagger}_n\cdot \partial_\tau \vec{\phi}^{\,2}_n)-p^2 \left(J_1|\vec{\phi}^{\,1\dagger}_n\cdot \vec{\phi}^{\,2}_n|^2+J_2|\vec{\phi}^{\, 2\dagger}_{n}\cdot \vec{\phi}_{n+1}^{\, 1}|^2\right)
    \right\}. 
\end{align}
Since the spin on the even sites belong to the conjugate representation, its Berry phase has the opposite sign compared with that of the spin on odd sites. 
We choose the ferromagnetic couplings, $J_1, J_2>0$, so that the classical vacua are given by the parallel spin fields.  
When $J_1=J_2$, we have the following symmetry, 
\begin{align}
    \vec{\phi}^{\,1}_n\mapsto \vec{\phi}^{\,2 *}_n,\quad 
    \vec{\phi}^{\,2}_n\mapsto \vec{\phi}^{\,1 *}_{n+1}, 
    \label{eq:oneunittrans}
\end{align}
and this becomes the charge conjugation symmetry in the infrared effective theory.

To derive the low-energy effective continuum theory, we denote $x=an$ by introducing the lattice constant $a$, and put
\begin{align}
    \vec{\phi}^{\, 1}_n=\vec{z}(x), \quad 
    \vec{\phi}^{\, 2}_n=\sqrt{1-|\vec{\ve}(x)|^2}\,\vec{z}(x)+\vec{\ve}(x), 
\end{align}
with 
\begin{align}
    \vec{z}^{\,\dagger} \cdot \vec{\ve}=0. 
\end{align}
When $\vec{\ve}=0$ and $\vec{z}$ is uniform, it describes the classical ground states. 
$\vec{\ve}$ describes the short-distance fluctuation and it turns out that $|\vec{\ve}\,|\sim O((\sqrt{(J_1+J_2)}\, p)^{-1})$. 
Especially when $|p|\gg 1$, we can integrate it out by the Gaussian integration to obtain the $2$d $\mathbb{C}P^{N-1}$ sigma model. 

When performing the Gaussian integral for $\vec{\ve}$, it is convenient to pay attention to the following point. As $|\vec{z}\,|^2=1$, we have 
\begin{equation}
    (\partial_\mu \vec{z}^{\,\dagger}) \cdot \vec{z}+\vec{z}^{\,\dagger}\cdot \partial_\mu z=0. 
\end{equation}
Due to the presence of the first term on the left-hand-side, $\partial_\mu \vec{z}$ does not belong to the space of $\vec{\ve}$. Still, we have
\begin{equation}
    \vec{z}^{\,\dagger} \cdot \mathrm{D}_\mu \vec{z}=0,
\end{equation}
where $\mathrm{D}_\mu \vec{z}=\partial_\mu \vec{z}-(\vec{z}^{\,\dagger} \cdot \partial_\mu \vec{z}\,)\vec{z}$, and thus $\mathrm{D}_\mu \vec{z}$ belongs to the space of $\vec{\ve}$. 
In particular, when we find $\vec{\ve}^{\,\dagger}\cdot \partial_\mu \vec{z}$, we may replace it by $\vec{\ve}^{\,\dagger}\cdot \mathrm{D}_\mu \vec{z}$. 

We now compute each term of the Euclidean action:
\begin{align}
    &\vec{\phi}^{\, 1 \dagger}_n\cdot\partial_\tau \vec{\phi}^{\,1}_n-\vec{\phi}^{\,2\dagger}_n\cdot\partial_\tau \vec{\phi}^{\, 2}_n
    =\vec{\ve}^{\,\dagger} \cdot \mathrm{D}_\tau \vec{z}-(\mathrm{D}_\tau \vec{z}^{\,\dagger})\cdot \vec{\ve},\\
    &|\vec{\phi}^{\, 1\dagger}_n\cdot \vec{\phi}^{\, 2}_n|^2
    =1-|\vec{\ve}\,|^2,\\
    &|\vec{\phi}^{\, 2\dagger}_n\cdot \vec{\phi}^{\,1}_{n+1}|^2
    =1-|\vec{\ve}\,|^2 +a (\vec{\ve}^{\,\dagger}\cdot \mathrm{D}_x \vec{z}+(\mathrm{D}_x\vec{z}^{\,\dagger})\cdot \vec{\ve}\,)-a^2 |\mathrm{D}_x\vec{z}\,|^2. 
\end{align}
The Euclidean Lagrangian then becomes 
\begin{align}
    \mathcal{L}&=
    p^2 (J_1+J_2)|\vec{\ve}\,|^2-p (\vec{\ve}^{\,\dagger}\cdot \mathrm{D}_\tau z-(\mathrm{D}_\tau \vec{z}^{\,\dagger})\cdot \ve)\nonumber\\
    &\qquad -p^2 J_2 a (\vec{\ve}^{\,\dagger}\cdot \mathrm{D}_x z+(\mathrm{D}_x \vec{z}^{\,\dagger})\cdot \ve) + p^2 J_2 a^2 |\mathrm{D}_x\vec{z}\,|^2.
\end{align}
Performing the Gaussian integration of $\vec{\ve}$, we find that 
\begin{align}
    \mathcal{L}_{\mathrm{eff}}& = p^2 J_2  a^2|\mathrm{D}_x\vec{z}\,|^2-\frac{p^2}{J_1+J_2}\left(J_2 a \mathrm{D}_x \vec{z}^{\,\dagger}-\frac{1}{p}\mathrm{D}_\tau \vec{z}^{\,\dagger}\right)\cdot\left(J_2 a \mathrm{D}_x \vec{z}+\frac{1}{p}\mathrm{D}_\tau \vec{z}\right)\\
    &=\frac{1}{J_1+J_2}|\mathrm{D}_\tau \vec{z}\,|^2+p^2 a^2\frac{J_1 J_2}{J_1+J_2}|\mathrm{D}_x\vec{z}\,|^2 + pa \frac{J_2}{J_1+J_2}\ve^{\mu\nu}\mathrm{D}_\mu\vec{z}^{\,\dagger}\cdot \mathrm{D}_\nu\vec{z}. 
\end{align}
Replacing $a\sum_n$ by the spatial integration $\int \diff x$, the Euclidean action is given by 
\begin{align}
    S=\int \diff \tau \int \diff x\left(\frac{1}{(J_1+J_2)a}|\mathrm{D}_\tau \vec{z}\,|^2+p^2 a\frac{J_1 J_2}{J_1+J_2}|\mathrm{D}_x\vec{z}\,|^2\right)+\im \theta\, Q_{\mathrm{top}},
\end{align}
with 
\begin{align}
    \theta=\pi p \left(1-\frac{J_1-J_2}{J_1+J_2}\right). 
\end{align}
By rescaling the imaginary time, $\tau\to \frac{1}{pa \sqrt{J_1J_2}}\tau$, this becomes the relativistic $\mathbb{C}P^{N-1}$ sigma model with $\frac{1}{g^2}=\frac{p \sqrt{J_1 J_2}}{J_1+J_2}$ and the above $\theta$ parameter.
Especially when $J_1=J_2$, $\theta\in \pi \mathbb{Z}$ and it enjoys the $C$ symmetry, which is the remnant of the one-unit lattice translation symmetry associated with the complex conjugation~\eqref{eq:oneunittrans}.

\bibliographystyle{utphys}
\bibliography{./QFT.bib,./refs.bib}
\end{document}